\documentclass[12pt]{article}
\usepackage{geometry}
\usepackage{graphicx} 
\usepackage{float}
\usepackage{amsmath}
\usepackage{authblk}
\usepackage{comment}
\usepackage{url}
\usepackage{xcolor}
\usepackage{rotating}  
\usepackage{booktabs}  
\usepackage{multirow}  

\title{Segregation and Ordering of Light Interstitials (B, C, H, and N) in Cr-Ni Alloys: Implications for Grain Boundary Stability in Superalloy Design}
\author[1,2]{Tyler D. Dole\v{z}al\footnote{corresponding authors: tyler.dolezal.1@us.af.mil; rodrigof@mit.edu; liju@mit.edu}}
\author[1]{Rodrigo Freitas$^*$}
\author[1,3]{Ju Li$^*$}
\affil[1]{Department of Materials Science and Engineering, Massachusetts Institute of Technology, Cambridge, MA, USA}
\affil[2]{Department of Engineering Physics, Air Force Institute of Technology, Wright-Patterson Air Force Base, OH, USA}
\affil[3]{Department of Nuclear Science and Engineering, Massachusetts Institute of Technology, Cambridge, MA, USA}

\begin{document}

\maketitle

\begin{abstract}
The segregation and ordering behavior of light interstitials (B, C, and N) in Cr\textsubscript{30}Ni is examined as these elements are critical for grain boundary (GB) stability and high-temperature mechanical performance in Ni-based superalloys. Using Monte Carlo simulations, we identify the chemical and structural preferences of these interstitials in both bulk and GB environments, aligning with experimental segregation and precipitation trends. Boron strongly prefers GBs over the bulk, where it enhances GB cohesion and stabilizes the GB structure. Uniquely, boron induces a structural transformation at higher concentrations, hinting at the formation of serrated GBs where boron content is high, which improves high-temperature mechanical performance. Carbon and nitrogen form carbide- and nitride-like motifs and exhibit limited GB solubility, reinforcing their precipitation tendencies. In support of ongoing hydrogen embrittlement mitigation strategies, we also examined hydrogen behavior. Hydrogen demonstrated chemical stability in the CrNi GB zone, suggesting it may preferentially migrate inward along Cr- and Ni-rich GBs while avoiding Mo-enriched regions, further supporting Mo’s role in mitigating embrittlement. These findings suggest that Mo-containing borides may serve as effective barriers against hydrogen-induced degradation by inhibiting H ingress and stabilizing GB cohesion. By elucidating the chemical and structural preferences of these light interstitials, this work provides a robust computational framework for guiding superalloy design toward improved high-temperature grain boundary stability, resistance to hydrogen embrittlement, and controlled chemical ordering.
\end{abstract}

\section{Introduction}
Nickel-based superalloys are indispensable for high-temperature applications due to their remarkable mechanical strength, thermal stability, and resistance to oxidation and creep. These properties are critical for demanding environments in aerospace propulsion, energy generation, and advanced manufacturing. However, several studies have highlighted limitations in these materials. For example, Hosseini and Popovich \cite{Hosseini2019} provided a comprehensive review of the mechanical properties of additively manufactured Inconel 718, identifying challenges such as porosity, anisotropy, and microstructural heterogeneity, which stem from the additive manufacturing process itself. These issues introduce residual stresses and defects that degrade tensile strength, fatigue life, and creep resistance. Jia and Gu \cite{Jia2014} demonstrated that selective laser melting (SLM) of Inconel 718 improves microstructure and densification when processed under optimal laser parameters, enhancing microhardness, wear resistance, and oxidation resistance through refined architectures and the formation of protective Cr\textsubscript{2}O\textsubscript{3} layers. Similarly, De Souza et al. \cite{desouzaWeldingAdditiveManufacturing2025} reviewed welding and additive manufacturing challenges in nickel-based superalloys, emphasizing hydrogen embrittlement (HE) as a significant issue. Their study detailed how welding parameters and post-weld treatments influence hydrogen uptake and subsequent embrittlement, particularly in the heat-affected zone.

In operational environments, prolonged exposure to extreme temperatures exacerbates material degradation. Hardy et al. \cite{hardySolvingRecentChallenges2020} highlighted the challenges of grain boundary weakening and coarsening of $\gamma '$ and $\gamma ''$ strengthening phases during extended high-temperature service, which significantly reduce creep resistance and tensile strength. Khalid and Mansoor \cite{khalidHydrogenEmbrittlementNickelBase2022} explored HE in Inconel 718, identifying weak sites such as $\delta$-phase interfaces and grain boundaries that are prone to decohesion and crack initiation. Their study emphasized the role of hydrogen-enhanced localized plasticity (HELP) and hydrogen-enhanced decohesion (HEDE) in accelerating embrittlement, with nanovoid nucleation at dislocation slip bands and intergranular decohesion identified as critical mechanisms. Feng et al. \cite{fengHydrogenEmbrittlementNiBased2024} investigated the impact of hydrogen on the fracture mechanism of wire arc additive-manufactured Inconel 625, demonstrating a transition from ductile microvoid coalescence to brittle quasi-cleavage, with preferential cracking observed at $\gamma$-matrix/Laves phase interfaces. While hydrogen enrichment weakened interface binding forces, the strong hydrogen capture ability of the Laves phase limited segregation at the interface. Baek et al. \cite{baekUltrasonicNanocrystalSurface2023} demonstrated that ultrasonic nanocrystal surface modification (UNSM) mitigates HE in additively manufactured Inconel 625 by refining surface grains, inducing compressive residual stresses, and reducing hydrogen diffusion, resulting in a 6.3\% improvement in elongation. Fu et al. \cite{fuEffectsHydrogenLoad2022} and Xu et al. \cite{xuHydrogenEmbrittlementBehavior2023} analyzed hydrogen’s impact on fatigue crack propagation, finding that it accelerates crack growth and alters microstructural deformation behavior. Lee et al. \cite{leeHydrogenassistedFailureInconel2022} investigated hydrogen-assisted failure in laser powder bed fused Inconel 718, demonstrating that direct aging (DA) exacerbates embrittlement due to intergranular cracking, while homogenization and aging (HA) mitigates hydrogen susceptibility by reducing trapping and promoting uniform distribution of strengthening phases. Recently, there has been growing interest in investigating the effects of light interstitial elements, such as boron, carbon, and nitrogen, on the properties of nickel-based superalloys.

Boron (B) plays a significant role in enhancing the performance and microstructural integrity of nickel-based superalloys. Antonov et al. \cite{antonovBoronTrappingDislocations2023} highlighted boron segregation at dislocations in additively manufactured superalloys, where it stabilizes dislocation networks and influences grain boundary properties. Kontis et al. \cite{kontisEffectBoronMechanical2014} showed that boron enhances creep resistance by forming $\gamma '$ layers at grain boundaries, which impede crack propagation and prevent the formation of continuous M\textsubscript{23}C\textsubscript{6} carbide films. Zhou et al. \cite{zhouRoleBoronConventional2008} demonstrated that boron improves stress-rupture life by facilitating dislocation transmission across grain boundaries, dissipating strain, and preventing premature crack initiation. Zhang et al. \cite{zhangMicrohardnessMicrostructureEvolution2016} demonstrated the effectiveness of TiB\textsubscript{2} reinforcements in Inconel 625, improving microhardness by forming Ti- and Mo-rich interfacial layers that enhance mechanical performance. Tian et al. \cite{tianSynergisticEffectsBoron2024} highlighted the synergistic effects of boron and rare earth elements (RE) in strengthening grain boundaries, improving plasticity, and reducing stress concentrations.

Carbon (C) also plays a vital role in strengthening nickel-based superalloys through carbide formation. Tekoglu et al. \cite{tekogluStrengtheningAdditivelyManufactured2023} demonstrated that nanocarbides formed during laser powder bed fusion improve yield and tensile strength by serving as dislocation barriers and enhancing grain refinement. Liu et al. \cite{liuTailoredMicrostructureEnhanced2025} optimized TiC/Inconel 718 composites using a dual-gradient printing strategy, achieving superior mechanical properties through tailored microstructure. Zhao et al. \cite{zhaoPrecipitationStabilityMicroproperty2018} investigated (Nb,Ti)C carbides in composite coatings, revealing their contribution to precipitation stability and hardness. Furthermore, Guan et al. \cite{guanMicrostructureMechanicalProperties2025} demonstrated that combining boron and carbon in TiB\textsubscript{2}-TiC phases significantly enhances mechanical properties, including microhardness and wear resistance, by refining grains and altering wear mechanisms.

Nitrogen (N) contributes to mechanical and tribological properties through nitride formation. Kim et al. \cite{kimStudyPittingCorrosion1996} showed that TiN coatings improve pitting corrosion resistance in Inconel 600, while Li et al. \cite{liInvestigatingInfluenceTiN2024} demonstrated that TiN reinforcements enhance the strength and wear resistance of Inconel 718. However, Lim et al. \cite{limEffectTitaniumNitride2023a} found that excessive TiN inclusions exacerbate brittle fracture and strain localization, reducing mechanical performance and underscoring the importance of inclusion control in advanced manufacturing processes.

Computational investigations into the behavior of light interstitials in nickel-based systems continue to provide valuable insights. Ji et al. \cite{jiEffectRefractoryElements2023} explored the influence of refractory elements (Re, W, Mo, Ta) on B diffusion in nickel-based single crystal superalloys, showing that these elements impede B diffusion. The study identified a trend of decreasing diffusion activation energy (Re $<$ Mo $<$ W $<$ Ta), attributed to differences in atomic radius and hybrid orbital interactions between B and the refractory elements. R\'{a}k et al. \cite{rakFirstprinciplesInvestigationBoron2014} examined the incorporation of B into nickel ferrite spinel (NiFe$_2$O$_4$) and revealed that B preferentially forms secondary phases such as B$_2$O$_3$, Fe$_3$BO$_5$, and Ni$_3$B$_2$O$_6$ rather than integrating into the spinel structure, due to the narrow stability domain of nickel ferrite and the high formation energies of B-related defects. Yang et al. \cite{yangFirstprinciplesInvestigationInteraction2016} focused on the adsorption, absorption, and diffusion of B on Ni (111) surfaces and in bulk Ni, finding that B preferentially adsorbs at HCP-hollow sites on the surface and occupies octahedral interstitial sites within the bulk, with a diffusion barrier of 1.65 eV between neighboring octahedral sites. In a follow-up study, Yang et al. \cite{yangFirstprinciplesInvestigationInteraction2017} investigated B behavior at grain boundaries (GBs) in Ni. They found that the $\Sigma3(111)[\bar{1}10]$ GB accelerates B diffusion but resists absorption due to its close-packed structure, while the $\Sigma5(210)[001]$ GB, with its larger cavities, acts as an effective trap for B, significantly hindering its diffusion. Yu et al. \cite{yuFirstPrinciplesStudyHydrogen2024} examined hydrogen solubility and its embrittlement effects at Ni/Cr\textsubscript{23}C\textsubscript{6} interfaces, finding that hydrogen accumulates at octahedral interstitial sites on the Ni side, reducing interface strength and increasing susceptibility to brittle fracture. Similarly, Hu et al. \cite{huSolubilityBoronCarbon2015} conducted a systematic study on the solubility and mobility of B, C, and N in various transition metals (TMs). They observed that interstitial formation energies follow a periodic trend governed by the interplay between unit cell volume and d-shell electron filling, with octahedral interstitial sites being energetically favorable for all three elements. David et al. \cite{davidFirstprinciplesStudyInsertion2020} further explored the solubility and diffusion of interstitial atoms (H, C, N, and O) in nickel. Their study confirmed that octahedral sites are the most stable positions, while tetrahedral sites exhibit higher elastic instability. Interestingly, C and N atoms showed additional stability at ``M'' sites, located between octahedral and tetrahedral positions, with temperature effects influencing solubility but having a minimal impact on migration energies.

This study advances the atomistic understanding of light interstitial behavior in nickel-based systems by examining the effects of B, C, H, and N at varying concentrations (1 at\% to 20 at\%). Unlike prior computational studies that primarily focused on pure nickel systems and limited dopant concentrations, this work investigates the Cr\textsubscript{30}Ni system, a representative model of a basic nickel superalloy composition, to explore the interplay between Ni and Cr in influencing interstitial behavior at a broad range of dopant levels. The analysis considers microstructural geometry and ordering in the pristine face-centered cubic (FCC) lattice as well as in the presence of GBs, offering a comprehensive view of interstitial effects in both bulk and defected environments. By systematically exploring multiple light interstitial types and concentrations within a binary alloy system, this work provides new insights into how compositional complexity and interstitial content influence structural stability and atomic-scale interactions.

\section{Methods}
\begin{figure}[H]
    \centering
    \includegraphics[width=\linewidth]{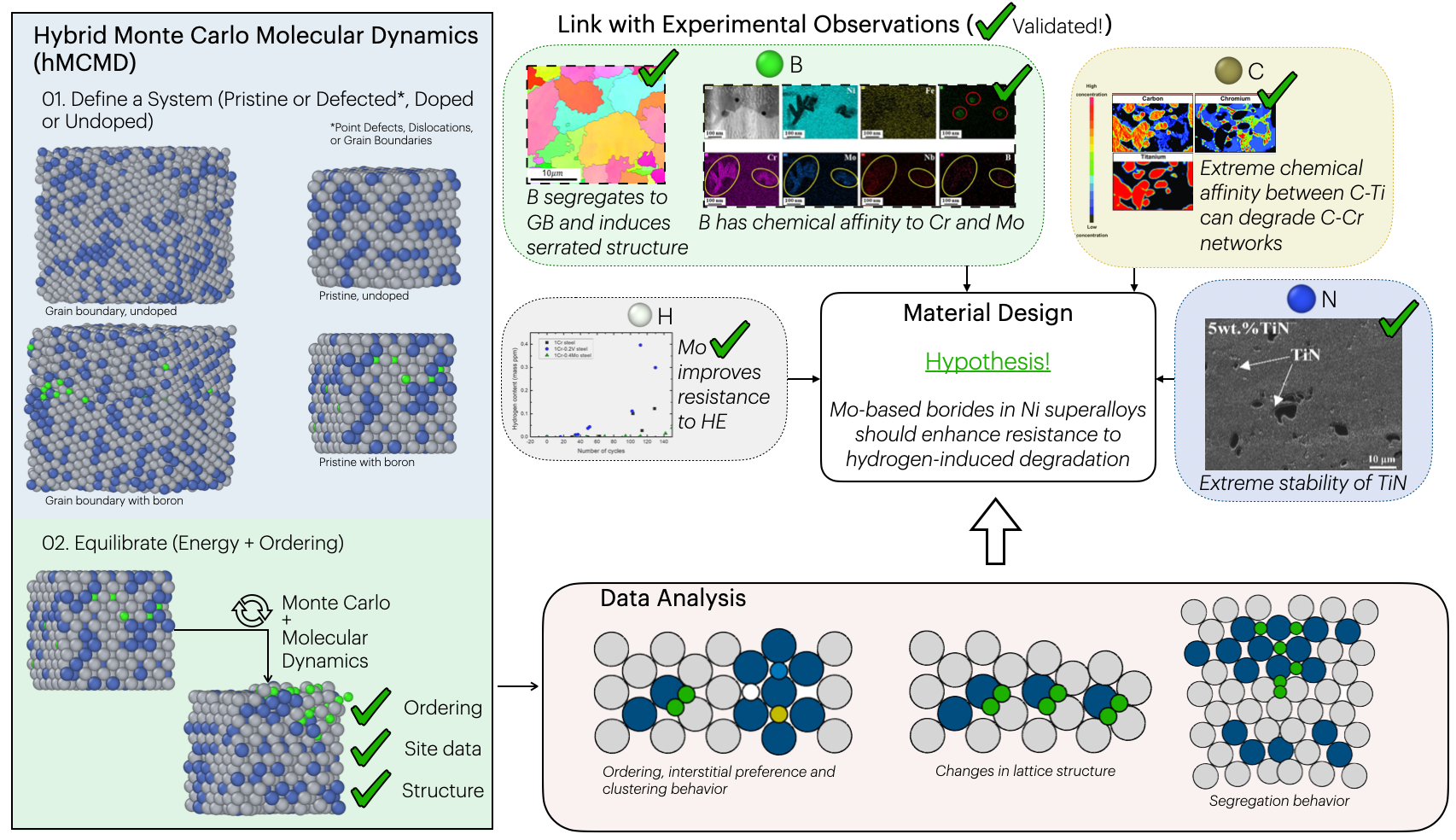}
    \caption{Schematic representation of the simulation workflow, some interesting topics that are explored, and experimental validation. The hybrid Monte Carlo Molecular Dynamics (hMCMD) approach is used to equilibrate atomic configurations and extract meaningful trends in ordering, interstitial site preference, and segregation behavior. The analysis phase includes clustering behavior and short-range order (SRO), and structural transformations. Several experimental observations are in agreement with our findings, such as boron segregation to grain boundaries, its chemical affinity for Cr and Mo, and how C-Ti interactions disrupt C-Cr networks. Experimental observations compiled from B \cite{tekogluMetalMatrixComposite2024b}, C \cite{chenAdditiveManufacturingTiC2024}, H \cite{leeComparativeStudyEffects2019}, and N \cite{liInvestigatingInfluenceTiN2024}. Finally, we propose a hypothesis that Mo-based borides in Ni superalloys could enhance resistance to hydrogen-induced degradation, inviting further experimental validation.}
    \label{fig:overview}
\end{figure}
\subsection{Preparing the Bulk Structures}
\subsubsection{Pristine Simulation Cells}
The bulk structures were initialized in an undoped Fm$\bar{3}$m cubic crystal system consisting of 6$\times$6$\times$6 conventional FCC unit cells, totaling 864 lattice positions. The supercell was compositionally randomized by distributing Ni and Cr atoms to achieve an overall composition of 30 atomic percent (at\%) Cr and 70 at\% Ni (Cr\textsubscript{30}Ni). Doped simulation cells were generated by randomly introducing light interstitials (B, C, H, or N) into the Cr\textsubscript{30}Ni supercell. The interstitial concentration was systematically varied from 1 at\% to 20 at\%, with increments of 2 at\% up to 10 at\%, followed by a direct increase to 20 at\% to evaluate the system near a more extreme doping concentration.
\subsubsection{Grain Boundary Simulation Cells}
To investigate the impact of GBs, a single GB was introduced into the center of a randomly populated pristine Cr\textsubscript{30}Ni FCC supercell consisting of 5,321 lattice positions. The GB simulation cell had dimensions of (39.52, 39.52, 39.52) \AA. The GB was introduced with a random orientation using Atomsk \cite{hirelAtomskToolManipulating2015} and fifteen distinct, undoped, GB simulation cells were generated and relaxed to determine the GB energy ($\gamma_{\rm GB}$). The GB energy was determined using Eq. \ref{eq:gb-eq}:
\begin{equation}\label{eq:gb-eq}
    \gamma_{\rm GB}(T=0) = \frac{E_{\rm GB} - N(E_{0}/N_{0})}{2A_{\rm GB}}\rm ,
\end{equation}
where $E_{\rm GB}$ is the energy of the GB simulation cell, $N$ is the number of atoms in the GB simulation cell, $E_0$ is the energy of the pristine structure, $N_0$ is the number of atoms in the pristine structure, and $A_{\rm GB}$ is the area of the GB (add a factor of 2 to account for 
a PBC GB). The complete relaxation scheme was a conjugate gradient (CG) structural relaxation with an energy and force convergence criteria of $1 \times 10^{-6}$ and $1 \times 10^{-8}$, respectively, followed by a canonical (NVT) molecular dynamics (MD) simulation for 5 ps at 1000 K, followed by CG structural relaxation with an energy and force convergence criteria of $1 \times 10^{-12}$. This was performed using the Large-scale Atomic/Molecular Massively Parallel Simulator (LAMMPS) software \cite{thompsonLAMMPSFlexibleSimulation2022a} and version 5.0.0 of the universal core neural network Preferred Potential (PFP) \cite{takamotoUniversalNeuralNetwork2022} with the D3 correction implemented through Matlantis \cite{Matlantis}. The configuration with the lowest $\gamma_{\rm GB}$ was selected for further analysis. The GB energy of the selected undoped Cr\textsubscript{30}Ni system was $\gamma_{\rm GB} = 0.800$ J/m\textsuperscript{2}, which is in good agreement with the range of experimentally measured and computationally predicted GB energies for Ni \cite{rohrerComparingCalculatedMeasured2010, sangidGrainBoundaryCharacterization2010}. Doped variations of the GB supercell were generated by randomly introducing light interstitials (B, C, H, or N) into the undoped GB structure up to 4 at\%.

\subsection{Computational Details}
The Monte Carlo (MC) simulations were performed on both the undoped and doped Cr\textsubscript{30}Ni systems, with five independent runs for each light interstitial concentration in the pristine systems. Each run was initialized from a randomized atomic configuration to ensure independent sampling and to minimize the risk of entrapment in local energy basins. Reported results include error bars representing the variability across the five MC runs. For the GB systems, only a single MC simulation was conducted per light interstitial concentration. This approach was informed by preliminary GB simulations which exhibited minimal statistical variability due to the significantly larger number of atoms and GB segregation in the GB simulation cells. Simulations were set to run until thorough equilibration was achieved in both the energy and short-range ordering of the systems. The MC algorithm trialed the following moves:
\begin{enumerate}
    \item \textbf{Swap atomic positions of metallic constituents:} Select two metallic atoms of different chemical types and attempt to swap their positions within the lattice.
    \item \textbf{Relocate a light interstitial atom near a new metallic host:} Select a light interstitial atom and attempt to place it near a different metallic atom within its nearest neighbor shell.
    \item \textbf{Introduce proximity between two light interstitial atoms:} Select two light interstitial atoms and attempt to place one of them within the first nearest neighbor shell of the other. If the first shell is fully occupied, the placement is attempted in the second nearest neighbor shell.
    \item \textbf{Separate two neighboring light interstitial atoms:} If a pair of neighboring light interstitials exists, select one and attempt to move it away from its nearest interstitial neighbor. This move balances the proximity adjustment, ensuring an opportunity to promote both segregation and aggregation of light interstitials.
    \item \textbf{Swap a metal neighbor of a light interstitial atom:} For a selected light interstitial atom, identify one of its nearest metallic neighbors and swap its position with that of another metal atom of a different chemical type.
\end{enumerate}
These moves were designed to explore the configurational space of the system effectively, balancing local aggregation and dispersion tendencies of the light interstitial atoms within the metallic matrix. Nearest neighbors were located within a cutoff radius of 2.75 \AA. The move was accepted or rejected based on the potential energy difference between the final and initial states, following the Metropolis criterion \cite{metropolisEquationStateCalculations1953}. The probability of acceptance is expressed as a Boltzmann probability:
\begin{equation}\label{eq:metropolis}
    P(\Delta U) = \min\{1, e^{-\Delta U/{k_{\rm B} T_{\rm sim}}}\}\rm ,
\end{equation}
where $\Delta U$ is the potential energy difference between the final and initial states, $k_\mathrm{B}$ is the Boltzmann constant in eV/K, and $T_{\rm sim}$ is the simulation temperature in K, which was set to 1073 K (800 \textsuperscript{o}C) for this work. For the MC simulations on the undoped Cr\textsubscript{30}Ni system, only the first move was trialed. The potential energy difference for each trial was calculated using LAMMPS, with structural relaxation performed via the CG method with an energy and force convergence criterion of $1 \times 10^{-12}$. The simulations were executed using version 5.0.0 of the universal core neural network PFP with the D3 correction implemented through Matlantis. Matlantis has been shown to be suitable for the modeling of Ni-based superalloys and their interactions with light interstitials \cite{mineComparisonMatlantisVASP2023, katoBoronCoordinationThreemembered2024a, choungRiseMachineLearning2024, hisamaMolecularDynamicsCatalystFree2024, hinumaNeuralNetworkPotential2024}. The MC-equilibrated GB structures underwent a final NVT MD simulation at 1073 K for 5 ps, followed by CG relaxation with energy and force convergence criteria of $1 \times 10^{-12}$.
\subsection{Analysis Methods}
\subsubsection{Atomic Ordering and Structure}
The Warren-Cowley short-range order (SRO) parameter ($\alpha_{ij}$) \cite{cowleyApproximateTheoryOrder1950, cowleyShortLongRangeOrder1960} was examined to quantify the degree of atomic ordering in the system. The SRO parameter is defined as:
\begin{equation}\label{eq:SRO}
    \alpha_{ij}(r) = 1 - \frac{P_{ij}(r)}{c_{j}} \rm ,
\end{equation}
where $P_{ij}(r)$ is the probability of finding atom $j$ as a neighbor to atom $i$ at a distance $r$, and $c_{j}$ is the atomic fraction of species $j$. This parameter was calculated for interstitial-metal (i-M) and metal-metal (M-M') pairs across varying light interstitial types (B, C, H, and N) and concentrations. The results were averaged over the five independent MC simulations for each system to improve statistical reliability. By analyzing $\alpha_{ij}$ as a function of light interstitial type and concentration, the study provides insight into the aggregation or segregation tendencies between alloy constituents, revealing how the presence and amount of light interstitials influence local atomic ordering.

Changes in the crystal structure were examined using polyhedral template matching (PTM) \cite{larsenRobustStructuralIdentification2016} as implemented in the OVITO Python API \cite{stukowskiVisualizationAnalysisAtomistic2009a}. This method allowed for the identification of local atomic environments by comparing the surrounding atomic arrangements to ideal polyhedral templates. For this analysis, light interstitial atoms were removed from the lattice to isolate and examine the spatial arrangement of the remaining metallic constituents. The PTM analysis was conducted as a function of the light interstitial type (B, C, H, or N) and concentration. This approach provides insights into how the presence and content of light interstitials affect the underlying crystal structure of the binary matrix.

\subsubsection{Excess Energy}

The excess energy ($E_{\text{excess}}$) was calculated as a function of light interstitial using Eq. \ref{eq:excess}:
\begin{equation}\label{eq:excess}
    E_{\text{excess}}^{(i)} = \frac{E^{(i)}_{\text{doped}} - N_{m}\left(\frac{E_0}{N_0}\right) - N_{i}E_{i}}{N_m + N_{i}} \rm ,
\end{equation}
where $E^{(i)}_{\text{doped}}$ is the averaged total energy of the simulation cell doped with interstitial species $i$, $N_{m}$ is the number of metal atoms in the doped simulation cell, $N_{i}$ is the number of light interstitials of species $i$ in the doped simulation cell, $E_{0}$ is the energy of the undoped simulation cell, and $N_0$ is the total number of atoms in the undoped cell. The term $E_{i}$ represents the chemical potential energy of the interstitial species derived from a reference state, which serves as a baseline for determining the relative energetic stability of the dopant within the alloy matrix. The excess energy is a measure of how favorable it is for B, C, H or N to remain in the MC equilibrated binary metal environment versus leaving to form a more preferred ordering.

The selection of reference states for interstitial chemical potential energy calculations is important in ensuring physically meaningful comparisons. Given that Ni-based superalloys commonly contain Ni, Cr, Fe, Mo, Nb and Ti, the energetic contributions of interstitial species (B, C, H, and N) were extracted from binary compounds with Ni, Cr, Fe, Mo, Nb and Ti. The compounds were compiled using Materials Project \cite{jainCommentaryMaterialsProject2013} and the compound with the lowest formation energy was used. The full data set is included in the Supplemental Materials (Tables \ref{tab:supplemental_materials_energies_1} and \ref{tab:supplemental_materials_energies_2}). To obtain $E_i$, the relaxed total energy of the reference state was computed, and the interstitial chemical potential energy per atom was determined from Eq.~\ref{eq:mu}:

\begin{equation}\label{eq:mu}
    E_{i} = \frac{E_{\text{comp}} - N_{m} \left(\frac{E_{\text{metal}}}{N_{\text{metal}}}\right)}{N_{i}}\text{ ,}
\end{equation}
where $E_{\text{comp}}$ is the total energy of the reference boride, carbide, hydride, or nitride compound, $E_{\text{metal}}$ is the total energy of the corresponding bulk metal system, and $N_{\text{metal}}$ is the number of metal atoms in the bulk metal system. In Eq.~\ref{eq:mu}, $N_{m}$ and $N_{i}$ represent the number of metal and interstitial atoms in the compound, respectively. The extracted interstitial energies and reference compound data are given in Table \ref{tab:interstitial_energies}. By using these reference states, the calculated $E_{\text{excess}}$ values more accurately capture the thermodynamic tendencies of B, C, H, and N in their preferred short-range ordered configurations. This approach ensures that $E_{\text{excess}}$ reflects the balance between interstitial dissolution and clustering within the alloy, while also enabling direct comparisons between different B-M, C-M, H-M, and N-M interactions. Importantly, the selected reference phases are not used to predict phase formation. Rather, they serve as consistent thermodynamic anchors to evaluate the relative chemical driving forces for interstitial ordering with various members of the Inconel metallic constituents.

Systematically comparing $E_{\text{excess}}$ across different reference compounds and interstitial types provides valuable insight into the thermodynamic preferences and chemical behavior of these species within the Ni-based superalloy environment. A negative $E_{\text{excess}}$ ($E_{\rm excess} < 0$) for a given interstitial with respect to a specific reference compound suggests that, within the chemically complex Cr-Ni matrix, the interstitial is thermodynamically stabilized through its finalized ordering. In contrast, a positive $E_{\text{excess}}$ ($E_{\rm excess} > 0$) indicates that, despite having access to its preferred local ordering within the MC-equilibrated Cr-Ni system, the interstitial would still be more stable in a region enriched in the reference metal species (e.g., Cr or Ti). For example, a positive $E_{\text{excess}}$ for B with respect to Cr-B ordering implies that B would prefer a more Cr-rich environment than is available in the Cr-Ni alloy matrix. Similarly, comparing $E_{\text{excess}}$ among B, C, H, and N enables evaluation of which interstitials are more likely to remain in solution versus contribute to chemical clustering. These calculations provide a thermodynamic framework for interpreting interstitial site preferences, segregation tendencies, and their potential influence on alloy microstructure and properties.

\subsubsection{Grain Boundary Energetics and Geometry}
The GB energy of the doped GB structures ($\gamma^{(\rm doped)}_{\rm GB}$) was computed according to Eq. \ref{eq:gb-eq} with the exception that it must account for the energy differences between the doped GB and pristine structures, i.e., $E_{\rm GB} \rightarrow E^{(\rm doped)}_{\rm GB}$ and $E_{0} \rightarrow E^{(\rm doped)}_{0}$. This analysis used the doped, MC equilibrated, structures to evaluate the stability of interstitial species (B, C, H, N) as a function of content. A decrease in $\gamma^{(\rm doped)}_{\text{GB}}$ with increasing interstitial content indicates a stabilization effect. Additionally, segregation energy ($E_{\text{seg}}$) for the different interstitials as a function of content was determined as defined in Eq. \ref{eq:segregation} \cite{lejcekGrainBoundarySegregation2010}:

\begin{equation}\label{eq:segregation}
    E^{(i)}_{\text{seg}} = \left( \frac{E^{(i)}_{\text{GB}}}{N_{\text{GB}}} - \frac{E^{(i)}_{\text{bulk}}}{N_{\text{bulk}}} \right) - \left( \frac{E_{\text{GB}}^{\text{undoped}}}{N_{\text{GB}}^{\text{undoped}}} - \frac{E_{\text{bulk}}^{\text{undoped}}}{N^{\text{undoped}}_{\text{bulk}}} \right) \rm ,
\end{equation}
where $ E^{(i)}_{\text{GB}} $ and $ N_{\text{GB}} $ represent the total energy and number of atoms in the GB system (doped with interstitial $i$), respectively. The terms $ E^{(i)}_{\text{bulk}} $ and $ N_{\text{bulk}} $ represent the total energy and number of atoms in the pristine system (doped with interstitial $i$), respectively. The terms $ E_{\text{GB}}^{\text{undoped}} $ and $ N_{\text{GB}}^{\text{undoped}} $ represent the total energy and number of atoms in the undoped GB system, respectively. The terms $ E_{\text{bulk}}^{\text{undoped}} $ and $ N_{\text{bulk}}^{\text{undoped}} $ represent the total energy and number of atoms in the undoped pristine system, respectively. This expression accounts for the energy difference between the doped and undoped GB systems, normalized per atom, and subtracts the corresponding energy difference for the pristine systems. The resulting $E_{\text{seg}}$ value indicates the relative stability of the light interstitial at the GB compared to its bulk solubility. Where $E_{\text{seg}} < 0$ indicates segregation is favorable and $E_{\text{seg}} > 0$ indicates segregation is unfavorable.

To assess changes in GB morphology, the relative free volume of metallic atoms (Cr and Ni) within the GB region was calculated as a function of interstitial type and concentration. For each metallic atom in the GB, the relative free volume ($\tilde{V}$) was defined as:
\begin{equation}\label{eq:free_volume}
    \tilde{V_j} = \frac{V_{j,\text{GB}} - \langle V_{\text{bulk}}\rangle}{\langle V_{\text{bulk}}\rangle} \textrm{ ,}
\end{equation}
where $V_{j,\text{GB}}$ is the atomic volume of metal atom $j$ in the GB, and $\langle V_{\text{bulk}}\rangle$ is the average atomic volume derived from the 5 equilibrated undoped pristine Cr\textsubscript{30}Ni structures. Using the compiled $\tilde{V}$ values, the mean ($\mu$) and standard deviation $(\sigma$) of the relative free volume across all metallic atoms within the GB zone was recorded. To quantify the smoothness of the GB plane, the root mean square (RMS) roughness ($R_q$) was computed using Eq.~\ref{eq:rmsr}:

\begin{equation}\label{eq:rmsr}
    R_q = \sqrt{\frac{1}{N_m}\sum_{j=1}^{N_m}(z_j - \bar{z})^2} \textrm{ ,}
\end{equation}
where $z_i$ is the height of the $j^{\text{th}}$ metallic atom, $\bar{z}$ is the mean height of atoms within the GB region, and $N_m$ is the total number of metallic atoms present in the GB zone. This quantity represents the degree of vertical variability or deviation of GB atomic positions from the average GB plane, with larger values of $R_q$ indicating greater surface roughness. To clearly illustrate how the light interstitials impact $R_q$, the absolute difference between the doped and undoped systems was taken, 
\begin{equation}\label{eq:abs_diff}
   \tilde{R}_q = R^{(i)}_q - R^{(\rm undoped)}_q \textrm{ ,}
\end{equation}
where negative values of $\tilde{R}_q$ represent a flatter GB plane and positive values correspond to increased roughness compared to the undoped GB plane. These geometric measures are summarized in Table~\ref{tab:free_volume}.

\section{Results and Discussion}
\subsection{Atomic Ordering and Structure}
\begin{figure}[H]
    \centering
    \includegraphics[width=0.8\linewidth]{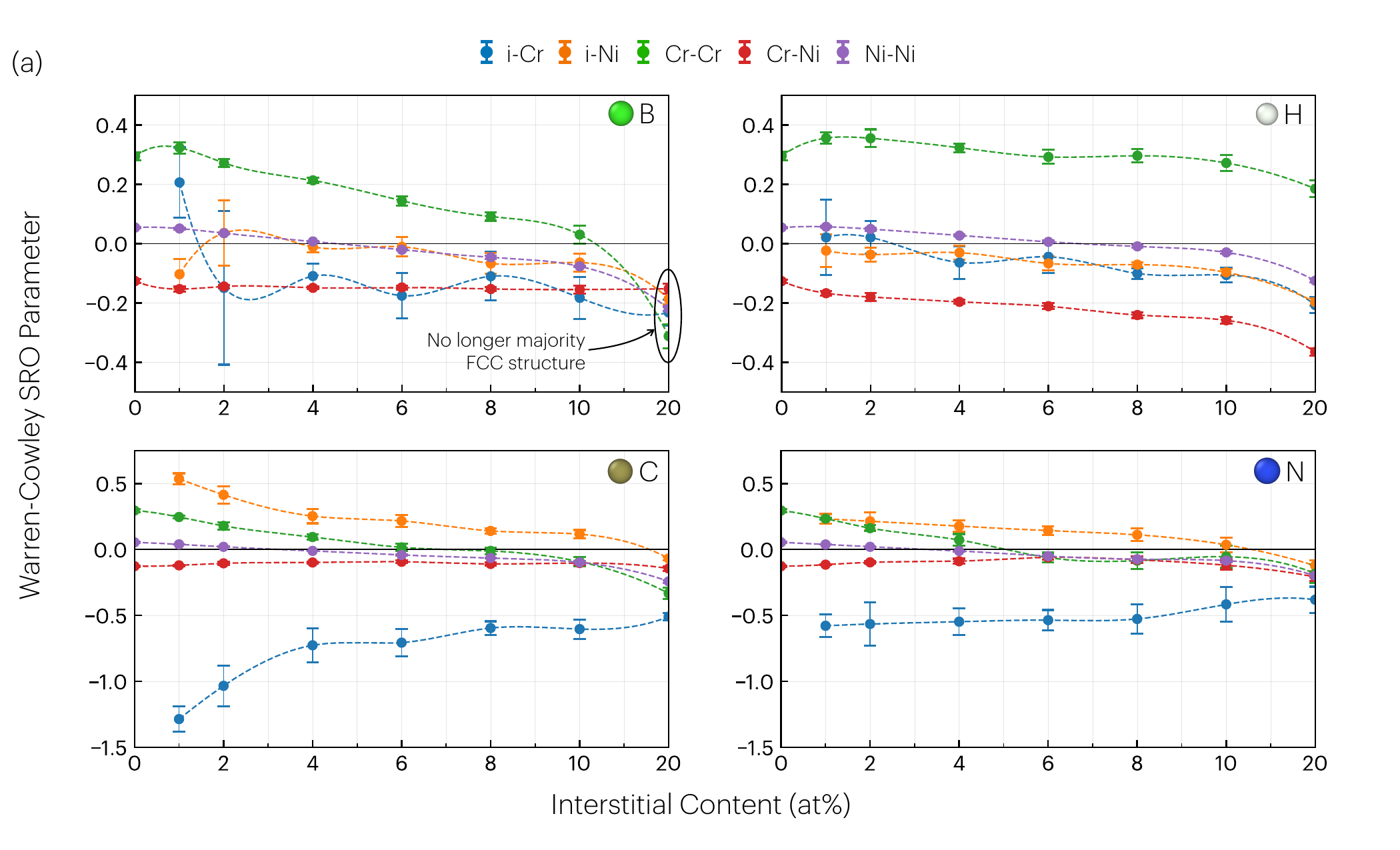}
    \includegraphics[width=0.8\linewidth]{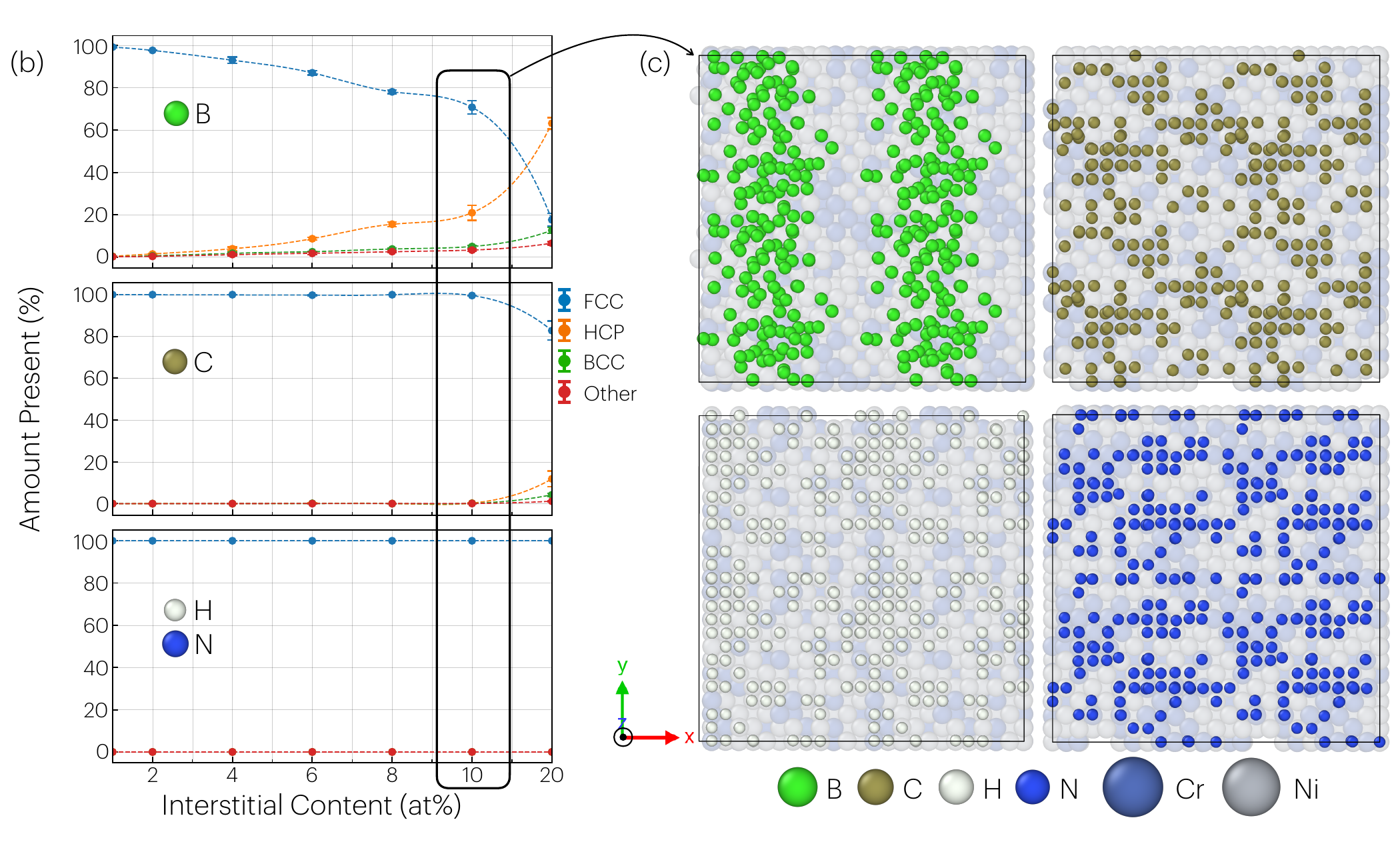}
    \caption{The (a) short-range order (SRO) parameter, as given in Eq. \ref{eq:SRO}, for the interstitial-metal (i-M) and metal-metal (M-M') pairs and (b) structure type in the pristine Cr\textsubscript{30}Ni supercell shown as a function of interstitial type and content. A single graph is shown for hydrogen or nitrogen, as their graphs were identical. (c) A final structure, replicated using a (2, 2, 1) supercell expansion to highlight clustering across the simulation cell boundaries.}
    \label{fig:sro_compiled}
\end{figure}

The Warren-Cowley SRO parameters ($\alpha_{ij}$), as defined in Eq.~\ref{eq:SRO}, are presented in Fig.~\ref{fig:sro_compiled}a for interstitial-Cr (blue), interstitial-Ni (orange), Cr-Cr (green), Cr-Ni (red), and Ni-Ni (purple) as a function of interstitial content for B, C, H, and N. These values were extracted from five independent MC equilibrated simulations, and the error bars represent the variability between these runs. To contextualize these ordering trends, the corresponding structural types are shown in Fig.~\ref{fig:sro_compiled}b, with cluster configurations provided in Fig.~\ref{fig:sro_compiled}c. Boron predominantly formed B\textsubscript{2} in the tetrahedral interstitial site, while C, H, and N preferentially occupied the octahedral interstitial site.

Across all interstitials, C and N exhibited the strongest affinity for Cr, as indicated by the consistently negative $\alpha_{\rm CCr}$ and $\alpha_{\rm NCr}$ values. However, their behavior diverged at higher concentrations due to differences in carbide and nitride motif formation. The $\alpha_{\rm CCr}$ value initially exhibited a strong negative trend but gradually increased with increasing C content, suggesting that C reached a saturation point with Cr, limiting further clustering. This behavior aligns with carbide stoichiometry, where a single C atom is typically coordinated by multiple Cr atoms (e.g., Cr$_{23}$C$_6$), leading to a depletion of available Cr neighbors at high C concentrations. In contrast, $\alpha_{\rm NCr}$ remained consistently negative across N content, indicating that N continued clustering with Cr without experiencing the same saturation constraints. This is likely due to the balanced Cr:N ratio in nitride formation (e.g., CN), allowing N to maintain strong Cr affinity even at higher concentrations. Meanwhile, B exhibited negative $\alpha_{\rm BCr}$ and $\alpha_{\rm BNi}$ values, with an affinity for Cr over Ni. The B-(Cr, Ni) interactions fluctuated significantly at 1 and 2 at\% B concentrations, but stabilized beyond 4~at\%, which coincided with the beginning of a structural transformation (Fig.~\ref{fig:sro_compiled}b). Unlike C, H, and N, which influenced local chemical ordering with little or no structural impact, B actively contributed to structural rearrangement. This effect was evident in the decreases in $\alpha_{\rm CrCr}$ and $\alpha_{\rm NiNi}$, indicating a structurally induced disruption to M-M' ordering. Similar to B, H showed a preference to cluster with both Cr and Ni, but more evenly. Because of this, H had a stronger effect on Cr-Ni interactions, reducing $\alpha_{\rm CrNi}$ more than any other interstitial, indicating that H modified Cr-Ni ordering more effectively. Overall, these trends highlight distinct interstitial behaviors on local lattice ordering: C and N exhibited the strongest affinity for Cr, but C reached a saturation limit, whereas N maintained clustering across all concentrations. B was the most disruptive to the overall structure, while both B and H had similar preferences on chemical ordering. These insights provide an interesting understanding of how different light interstitials modify Cr-Ni systems, with implications for chemical stability and early-stage microstructural evolution.

Recent experimental work~\cite{niuManipulatingInterstitialCarbon2020} examined the impact of C and H in the octahedral interstitial site on Ni microstructure. Their findings confirmed that both C and H preferentially occupy octahedral sites, supporting the site preferences observed in this study. Additionally, they reported that interstitial C inhibited unwanted hydrogenation, further highlighting its role in modifying Ni’s chemical environment. This aligns with the well-established tendency of Cr and C to form large-scale carbide networks with complex morphologies~\cite{wangInsightLowCycle2023, zhangSynergyPhaseMC2024, liInfluenceCarbidesPores2024}, further reinforcing the observed preference for extensive C-Cr clustering. 

Recent first-principles calculations support site preference and ordering observed here, demonstrating that C and N favor octahedral interstitial sites in FCC metals, whereas B exhibits different behavior, with a higher formation energy and migration barrier \cite{huSolubilityBoronCarbon2015}, which may contribute to its localized clustering behavior. Additionally, the solubility and migration behavior of these elements follow periodic trends, with C exhibiting lower formation energies and higher mobility compared to B and N, potentially driving its stronger clustering tendencies. Further insight into interstitial site preferences and diffusion mechanisms is provided by density functional theory (DFT) calculations \cite{davidFirstprinciplesStudyInsertion2020} which confirmed that C and N preferentially occupy octahedral sites. The charge transfer analysis showed that C and N strongly interacted with neighboring Ni atoms, while H exhibited weak Ni interactions, supporting its reduced clustering tendency.

The shared ordering observed between C and N with Cr suggests the potential for synergistic effects, as N has been shown to promote the formation of carbonitrides \cite{nabaviMetallurgicalEffectsNitrogen2019}. Given that both C and N exhibited strong clustering with Cr in these simulations, their co-presence in a system could enhance carbonitride formation. This adds an additional complexity to the idea that interstitial-driven ordering in Ni-based alloys is governed not only by individual solute behaviors but also by their collective interactions with metallic elements. Lastly, B was unique in that it was the only light interstitial to drive a structural transition (see Fig. \ref{fig:sro_compiled}b). This is particularly notable given B’s well-known tendency to segregate to GBs and strengthen them by altering GB morphology through induced serrations \cite{kontisEffectBoronMechanical2014, koulMechanismSerratedGrain1983, danflouFormationSerratedGrain1992, kontisRoleBoronImproving2017, tekogluMetalMatrixComposite2024b}. Unlike C, N, and H, which maintained the FCC lattice even at high concentrations, B progressively destabilized the FCC phase, ultimately leading to a majority transformation to HCP at 20~at\% B content. This drastic structural shift underscores a fundamental difference in how B interacts with the Ni-based matrix compared to other interstitials. The FCC-to-HCP transition supports the common observation that, in regions of high B concentration such as GBs, B can promote local structural transformations. Further investigations were performed on the deformed Cr\textsubscript{30}Ni B-doped structures. When B atoms were removed and the structures were re-relaxed, the FCC phase was largely restored, with more than 95\% of the system reverting to its original structure. However, small perturbations remained, with trace amounts of HCP and `Other' structural deviations, indicating that while B-induced lattice disruptions are mostly reversible, residual irregularities persist. This suggests that the destabilization arises from a B-driven perturbation rather than a permanent phase transformation. The reversibility highlights the transient nature of B-induced structural disruptions, reinforcing the idea that B acts as a local structure modifier rather than an agent of permanent phase change.
\subsection{Excess Energy}
\begin{figure}[H]
    \centering
    \includegraphics[width=\linewidth]{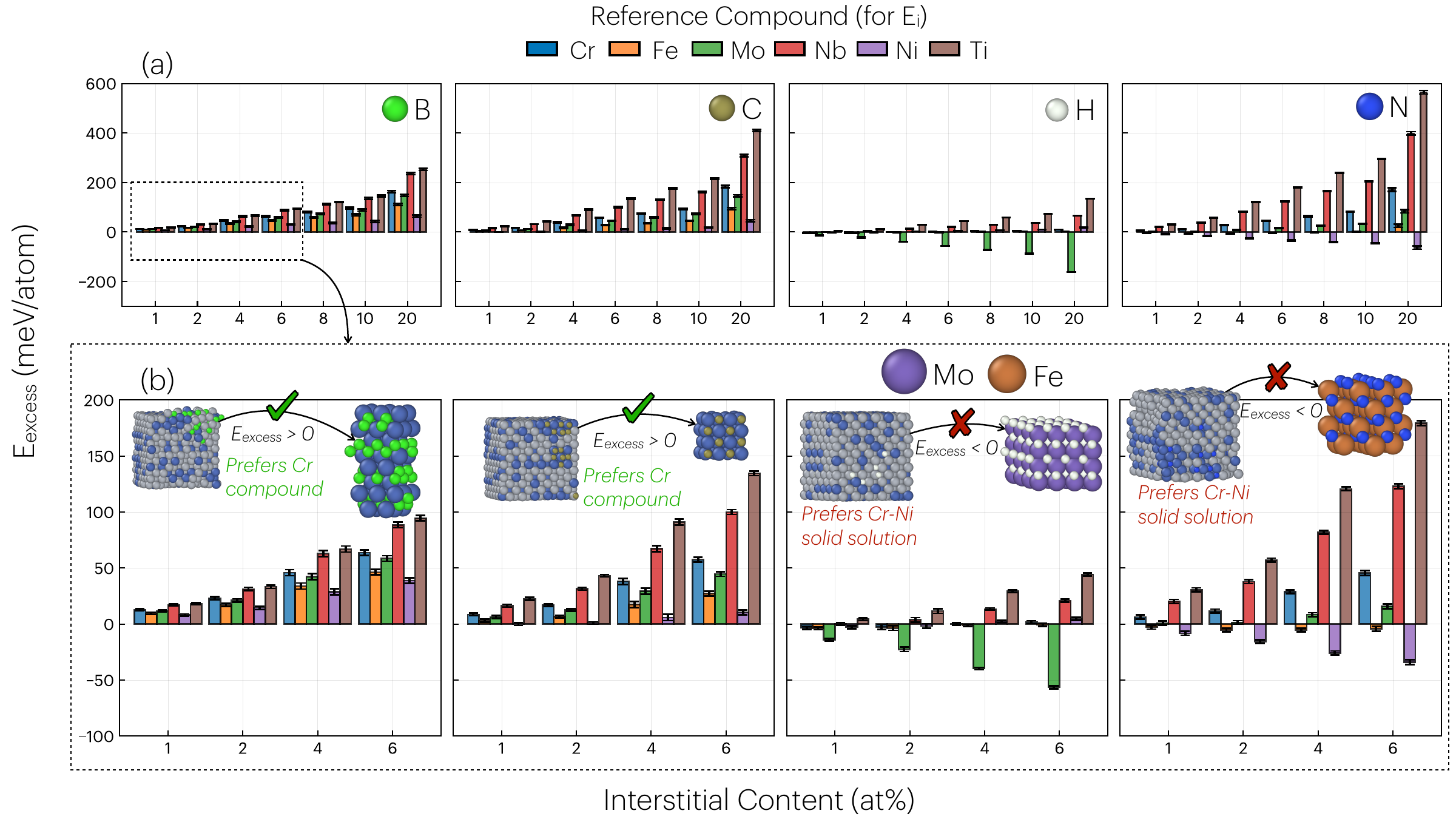}
    \caption{(a) Excess energy ($E_{\text{excess}}$), as defined in Eq.~\ref{eq:excess}, plotted against interstitial type, concentration, and the reference compound used to calculate $E_{i}$ (Eq.~\ref{eq:mu}). (b) A magnified view of $E_{\text{excess}}$ for interstitial concentrations of 6~at\% or lower, with schematics illustrating key trends and behaviors that $E_{\text{excess}}$ helps reveal. The label corresponds to the metal component of each reference compound, e.g., Cr for Cr–i (i = B, C, H, N) or Fe for Fe–i (i = B, C, H, N), etc.}
    \label{fig:excess_energy}
\end{figure}

Figure~\ref{fig:excess_energy} presents the calculated excess energy, $E_{\text{excess}}$, for B, C, H, and N as a function of interstitial content. The color of each bar corresponds to the reference compound used to extract $E_{i}$, which has been compiled in Table \ref{tab:interstitial_energies}. As a reminder, a more positive $E_{\text{excess}}$ value indicates a preference for that interstitial to segregate and form the compound/ordering associated with the reference structure, while negative values suggest a tendency to remain within the mixed-metal matrix. Importantly, because the host lattice was equilibrated through MC simulations prior to these calculations, there is a confidence that each light interstitial has reached its preferred ordering and lowest-energy configuration within the Cr\textsubscript{30}Ni system before being compared to reference compounds.

\begin{table}[H]
    \centering
    \renewcommand{\arraystretch}{1.0} 
    \caption{Reference compound data and computed interstitial chemical potential energies ($E_i$) as defined in Eq. \ref{eq:mu}.}
    \label{tab:interstitial_energies}
    \resizebox{\linewidth}{!}{ 
    \begin{tabular}{c c c}
        \toprule\toprule
        & \textbf{Boron (B)} & \textbf{Carbon (C)} \\
        \midrule
        \multirow{5}{*}{\rotatebox{90}{}}
        & \begin{tabular}{l l c}
            Compound & Space Group & $E_i$ (eV/atom) \\
            \midrule
            CrB   & I4$_{1}$/amd & -7.619 \\
            FeB   & Cmcm & -7.313 \\
            MoB   & I4$_1$/amd & -7.525 \\
            NbB   & Cmcm & -8.054 \\
            NiB   & Cmcm & -7.026 \\
            TiB$_2$   & P6/mmm & -8.159 \\
        \end{tabular}
        & \begin{tabular}{l l c}
            Compound & Space Group & $E_i$ (eV/atom) \\
            \midrule
            Cr$_{23}$C$_6$   & Fm$\bar{3}$m & -8.604 \\
            Fe$_{23}$C$_6$   & Fm$\bar{3}$m & -8.073 \\
            Mo$_2$C   & Pbcn & -8.380 \\
            Nb$_2$C   & Pnma & -9.354 \\
            Ni$_3$C   & R$\bar{3}$c & -7.776 \\
            Ti$_2$C   & Fd$\bar{3}$m & -9.963 \\
        \end{tabular} \\
        \midrule
        & \textbf{Hydrogen (H)} & \textbf{Nitrogen (N)} \\
        \midrule
        \multirow{5}{*}{\rotatebox{90}{}}
        & \begin{tabular}{l l c}
            Compound & Space Group & $E_i$ (eV/atom) \\
            \midrule
            CrH   & Fm$\bar{3}$m & -2.380 \\
            FeH   & Fm$\bar{3}$m & -2.339 \\
            MoH   & P6$_3$/mmc & -1.355 \\
            NbH$_2$   & Fm$\bar{3}$m & -2.720 \\
            NiH   & Fm$\bar{3}$m & -2.434 \\
            TiH$_2$   & I4/mmm & -3.127 \\
        \end{tabular}
        & \begin{tabular}{l l c}
            Compound & Space Group & $E_i$ (eV/atom) \\
            \midrule
            Cr$_2$N   & P$\bar{3}$1m & -6.645 \\
            Fe$_3$N   & P6$_3$22 & -5.769 \\
            MoN   & P6$_3$mc & -6.122 \\
            Nb$_2$N   & P$\bar{3}$1m & -8.007 \\
            Ni$_3$N   & P6$_3$22 & -5.241 \\
            TiN   & Fm$\bar{3}$m & -9.004 \\
        \end{tabular} \\
        \bottomrule\bottomrule
    \end{tabular}
    }
\end{table}

The most immediate and striking observation is that all light interstitials exhibit a strong preference to segregate with Ti, with C and N showing the highest $E_{\text{excess}}$ values. This trend aligns with the well-documented affinity of these elements for Ti, known to form stable borides (TiB\textsubscript{2}), carbides (e.g., TiC), hydrides (TiH\textsubscript{2}) and nitrides (e.g., TiN) \cite{philippStabilityTitaniumNitride1961, philippChemicalReactionsCarbides1966, piersonHandbookRefractoryCarbides1996, guTransitionMetalBorides2008, setoyamaMechanicalPropertiesTitanium2004, luoFormationTitaniumHydride2006, xuMechanicalPropertiesTitanium2007, friedrichSynthesisBinaryTransition2011}. Another key observation is that B exhibits positive excess energy values across all reference structures, even at the lowest concentrations, indicating negligible solubility in the Cr\textsubscript{30}Ni lattice and a strong chemical affinity for compound-forming elements, as reflected in positive $E_{\text{excess}}$ values. This result is in excellent agreement with experimental data for B in both pure Ni and pure Cr at 800\,$^\circ$C, where B readily forms Ni\textsubscript{3}B and Cr\textsubscript{2}B~\cite{portnoiPhaseDiagramSystem1967a, liaoBCrBoronChromiumSystem1986, oikawaExperimentalInvestigationThermodynamic2022}. The presence of Cr does not improve the solubility of B in the Ni matrix, further reinforcing its strong tendency toward segregation and precipitation.

In a typical Ni-based superalloy, such as Inconels, the composition consists primarily of Ni, followed by Cr, Mo, Fe, and small amounts of Ti. If each reference compound in this analysis represents a potential ``segregation route'' for B, then multiple pathways were available for its escape from the Cr\textsubscript{30}Ni matrix. Based on the magnitude of $E_{\text{excess}}$, TiB\textsubscript{2} was the most favorable segregation route. However, in a realistic alloy system, the probability of B aggregating enough Ti to precipitate as TiB\textsubscript{2} is low. Instead, B is more likely to encounter Cr, Mo, Fe, and some Nb where it showed a stronger preference for Mo and Nb over Cr and Fe. Thus, if B is introduced into a NiCrMoFeTi mixed-metal system, it should preferentially segregate in regions rich in Nb and Mo rather than in the Ni $\gamma$ matrix. Given B’s known tendency to segregate to GBs, this hypothesis can be refined further: B should be found at GBs where it would aggregate with Nb and Mo. Experimental validation of this hypothesis can be found in Ref.~\cite{tekogluMetalMatrixComposite2024b,tekogluSuperiorHightemperatureMechanical2024c}, which reported ordering of Cr, Mo, Nb, and B in Inconel 625 and Cr, Mo, Nb, and B in Inconel 718, respectively. In these studies, B was introduced via TiB\textsubscript{2} and ZrB\textsubscript{2} doping. In the absence of sufficient Ti, B dissociated and re-ordered with Cr, Mo, and Nb aligning with the $E_{\text{excess}}$ analysis.

Recent experimental studies on the partitioning behavior of Mo and Cr in Ni-based superalloys have shown that Cr preferentially partitions to the $\gamma$ matrix, while Mo’s distribution shifts dynamically depending on Cr content \cite{wangNewInsightsPartitioning2025, liuEffectMoAddition2015, wangInfluenceReCr2016, maPartitioningBehaviorLattice2021}. It was determined that higher Cr concentrations reduced Mo’s solubility in $\gamma'$ precipitates, driving its redistribution to the $\gamma$ phase. This redistribution behavior suggests a complex interplay between Cr and Mo, where Cr-rich environments indirectly influence the stability and partitioning tendencies of Mo within the alloy.  Given this interplay, the addition of B introduces another layer of complexity. Specifically, B-enriched regions should also be Cr/Mo-rich. Since Mo’s solubility in $\gamma'$ decreases as Cr content rises, B segregation could further reinforce Mo partitioning to the $\gamma$ matrix, altering the Cr/Mo balance. This suggests that B’s presence at GBs may not only influence segregation behavior (and morphology) but also stabilize Cr/Mo-rich GB phases while further depleting Mo from $\gamma'$. The interdependence between B segregation and Cr/Mo redistribution highlights a potential avenue for tailoring microstructural properties, warranting further experimental validation.

Transitioning now to C, the excess energy results in Fig.~\ref{fig:excess_energy}b highlights its limited solubility in the Cr\textsubscript{30}Ni matrix. Compared to pure Ni and pure Cr, where C exhibits practically no solubility at 800\,$^\circ$C~\cite{singletonCNiCarbonNickelSystem1989, venkatramanCCrCarbonChromiumSystem1990}, the addition of Cr has no measurable effect on stabilizing C in solid solution, as expected. This is reflected in the positive $E_{\text{excess}}$ values observed across all dopant concentrations, indicating a consistent thermodynamic drive for segregation and compound formation. The negative and low $E_{\text{excess}}$ values for NiC suggests that C would remained preferentially clustered with Cr as an interstitial in the Cr-Ni lattice rather than form a Ni-based carbide. It is particularly interesting to observe positive $E_{\text{excess}}$ values for the other alloy constituents, with C-Ti and C-Nb displaying the largest positive value. The positive values indicate that these elements serve as favorable segregation routes, despite C achieving strong ordering with Cr in the equilibrated Cr\textsubscript{30}Ni cell. For Mo, this observation aligns well with ab initio calculations reporting the strong thermodynamic stability of Mo-C \cite{liInitioThermodynamicStability2022}, as well as experimental work demonstrating the persistence of Mo-based carbides in Ni/Fe Cr-Mo systems \cite{xiePrecipitationStrengtheningBehavior2008, cabrolExperimentalInvestigationThermodynamic2013}. Additionally, it supports the recent observations of both Cr and Mo participating in carbide formation at GBs \cite{bianGrainBoundaryDiffusion2024} in Ni-Cr-Mo alloys. While Fe-based carbides are energetically favorable, they are not commonly observed in Ni-based systems due to Fe’s solubility in the $\gamma$ matrix. The positive $E_{\text{excess}}$ value suggests that Fe-C formation would be preferable over Cr-C \textit{if} Fe were available for carbide formation. However, Fe remains largely stable in solid solution and lacks a significant segregation drive, preventing Fe-based carbides from forming in appreciable amounts. 

The chemical affinity between C and Ti is significant considering experimental evidence which supports the competitive interaction between TiC and Cr-based carbides \cite{weiMicrostructureEvolutionCreeprupture2023, chenAdditiveManufacturingTiC2024}. The study from Ref.~\cite{chenAdditiveManufacturingTiC2024} confirmed that TiC exhibits strong thermodynamic stability and a high melting point, often outcompeting other carbides in metal matrices. In systems containing Cr, TiC influenced carbide formation by altering the solubility and stability of Cr-rich carbides \cite{weiMicrostructureEvolutionCreeprupture2023, chenAdditiveManufacturingTiC2024}, consistent with the fact that TiC had the highest $E_{\text{excess}}$ among all reference carbides.  Additionally, the redistribution of C in TiC-containing systems was shown to impact the dissolution and re-precipitation of Cr$_{23}$C$_6$ carbides \cite{chenAdditiveManufacturingTiC2024} which can weaken GB stability \cite{weiMicrostructureEvolutionCreeprupture2023}. Together, these findings reinforce the critical role of Ti in disrupting Cr-based carbide stability and highlight its broader implications in Ni-based alloys. The competition between Cr- and Ti-based carbides presents a significant challenge for microstructural stability, especially when doping with TiC \cite{tekogluMetalMatrixComposite2024b, chenAdditiveManufacturingTiC2024, weiMicrostructureEvolutionCreeprupture2023}, emphasizing the need for careful control of Ti content to mitigate adverse effects on GB cohesion and mechanical performance.

Given the strong affinity between C-Nb, it is important to note how NbC plays a crucial role in carbide formation and GB stabilization in Ni-based alloys. Unlike TiC, which disrupts Cr-rich carbides, NbC primarily increases the overall carbide fraction, forming both within grains and at GBs \cite{liuEffectsNbAddition2024}. The work of Ref. \cite{liuEffectsNbAddition2024} found that Nb-based carbides serve as a strong pinning agent, reducing GB mobility and refining the microstructure, thereby enhancing creep resistance and tensile strength. The work of Li et al. \cite{liEffectAnnealingTemperature2024} discovered that, as annealing temperature increases, NbC stability decreased relative to Cr-Mo carbides, which lead to a transition in carbide composition from (Nb, Ti)C to (Cr, Mo, W)C. This suggests that NbC and TiC play a dynamic role in carbide phase evolution, influencing long-term microstructural stability in Ni-based alloys. These results underscore the importance of leveraging atomistic insights to capture the fundamental mechanisms governing metal stability in the presence of light interstitials, providing a more detailed understanding of how elemental interactions at the atomic scale influence larger-scale behavior.

Hydrogen is a unique light interstitial to consider due to its well-documented role in degrading mechanical performance. Understanding its preferred ordering with the metallic constituents of Ni-based superalloys may provide atomistic insights into strategies for mitigating deleterious H aggregation within the microstructure. What stands out most with H is its clear avoidance of Mo and its near-neutral interactions with Ni, Cr, and Fe. This behavior suggests that while H does not exhibit a strong driving force to cluster with major alloying elements, its selective repulsion from Mo may have implications for H trapping mechanisms and embrittlement resistance. Recently, experimental studies have demonstrated that Mo-carbides exhibit superior H-trapping capabilities, significantly enhancing resistance to HE \cite{leeRoleMoCarbides2015, leeEffectsVanadiumCarbides2016, leeComparativeStudyEffects2019}. First-principles calculations further confirm that Mo reduces H solubility in metallic systems, offering greater HE resistance than W \cite{liuEffectsMoConcentration2024}. Moreover, Mo$_2$C carbides have been experimentally shown to serve as effective H traps, lowering H diffusivity and improving HE resistance \cite{eskinjaInfluenceMoCarbides2024}. Additionally, Ref.~\cite{kimHydrogenEmbrittlementBehavior2025} demonstrated that Mo-containing carbides improved HE resistance by stabilizing GBs and mitigating H-assisted fracture.  These experimental findings align with the segregation trends which indicated repulsion of H from Mo, reinforcing the role of Mo-rich regions as effective H diffusion barriers, potentially reducing embrittlement risks in Ni-based alloys. This highlights the critical role of Mo in both carbide formation and microstructural stability under H exposure, further underscoring its importance in alloy design.

Interestingly, recent experimental work has demonstrated that B, like Mo, enhances HE resistance by segregating to GBs and repelling H \cite{hachetSegregationPriorAustenite2024}. Thermal Desorption Spectroscopy (TDS) measurements indicated that B passivates GBs against H accumulation. Ab initio calculations \cite{hachetSegregationPriorAustenite2024} further confirmed that B has a stronger attraction to GBs than H, suggesting a repulsive interaction that prevents H segregation. Given B’s well-documented co-segregation with Mo and Cr at GBs, this raises the possibility of a synergistic effect in Ni-based alloys, where B and Mo together could provide an effective barrier against H embrittlement. The $E_{\text{excess}}$ analysis provides intriguing support for this hypothesis, as B exhibited a strong segregation tendency with Mo. This suggests that, in Ni-based systems, GBs enriched in B and Mo could form highly resistant interfaces, where Mo-based borides play a critical role in repelling H and collectively mitigating HE.  If such an effect extends to B-Mo-rich GBs in Inconel alloys, it would present a new alloy design strategy for improving H resistance. Specifically, the observed energetic preference for B and Mo to segregate together implies that engineering GB compositions to promote MoB boride formation could create self-stabilizing, H-resistant interfaces, preventing embrittlement and extending alloy performance in H-exposed environments.

Attention now shifts to N, the final light interstitial. Similar to C, N exhibited no solubility in the Ni lattice even in the presence of Cr, consistent with experimental data and phase equilibria studies~\cite{kowandaSolubilityNitrogenLiquid2003, liNitrogenSolubilityMolten2023, ivanchenkoPhaseEquilibriaStability1996}. These results further confirm the limited capacity of the Cr\textsubscript{30}Ni matrix to accommodate N in solid solution under the simulated conditions. Like the other light interstitials, N showed a strong preference for Nb and Ti. The shared affinity of C and N with Nb provides further support to the hypothesis that Nb competes in carbide and nitride formation in Ni superalloys \cite{smithRoleNiobiumWrought2005}. Referencing Fig. \ref{fig:excess_energy}a, N clearly demonstrated an extreme chemical affinity for Ti, with significantly higher $E_{\text{excess}}$ values than any other light interstitial. This suggests that TiN formation is not only thermodynamically favorable but also a dominant segregation pathway for N, far outweighing its interactions with Cr or Mo. Experimental observations strongly support this trend, with TiN inclusions readily forming in Ni-based alloys and playing a critical role in nitrogen stabilization \cite{liInvestigatingInfluenceTiN2024, limEffectTitaniumNitride2023a}. Both studies demonstrated that TiN remains thermodynamically stable under extreme processing conditions, reinforcing that Ti-N bonding is one of the strongest interstitial-metal interactions in Ni-based systems. However, while TiN formation is energetically favorable, its impact on mechanical properties must be carefully considered. The incorporation of TiN has been shown to enhance tensile strength and wear resistance, yet excessive TiN content leads to particle agglomeration, increasing embrittlement and reducing ductility \cite{liInvestigatingInfluenceTiN2024}. Additionally, TiN was found to modify GB chemistry by reducing Laves phase formation and promoting a Nb-Mo diffusion layer at the TiN/IN718 interface. This is particularly noteworthy given that $E_{\text{excess}}$ analysis indicated an affinity between N and Mo. Given the extreme chemical affinity between Ti and N, TiN precipitation is virtually unavoidable in N-bearing Ni-based alloys. These findings underscore the need for precise N content control in alloy design, especially in additively manufactured Inconel systems, where unintended N incorporation could lead to excessive TiN formation and associated mechanical degradation.

\subsection{Grain Boundary Analysis}
\begin{figure}[H]
    \centering
    \includegraphics[width=\linewidth]{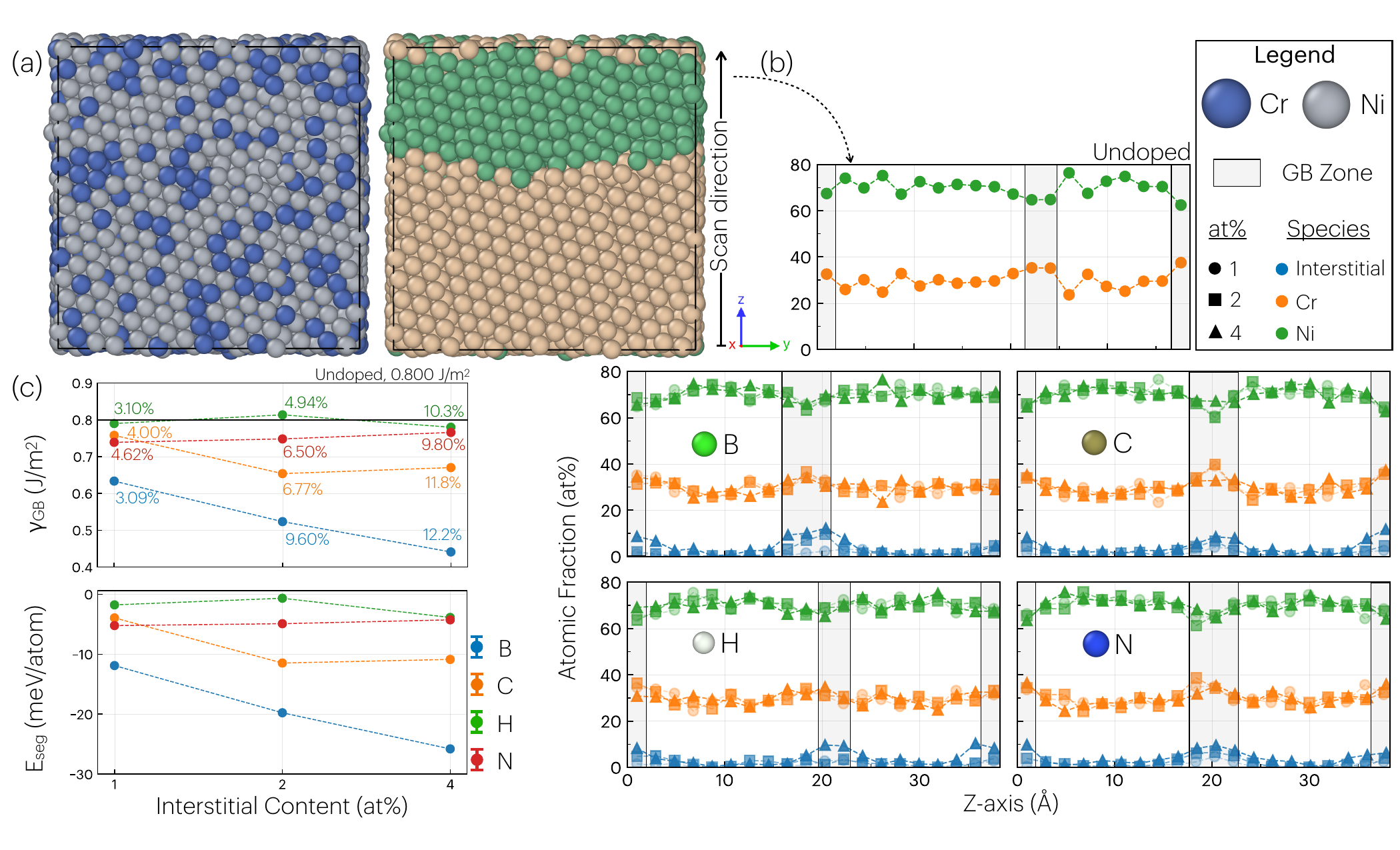}
    \includegraphics[width=\linewidth]{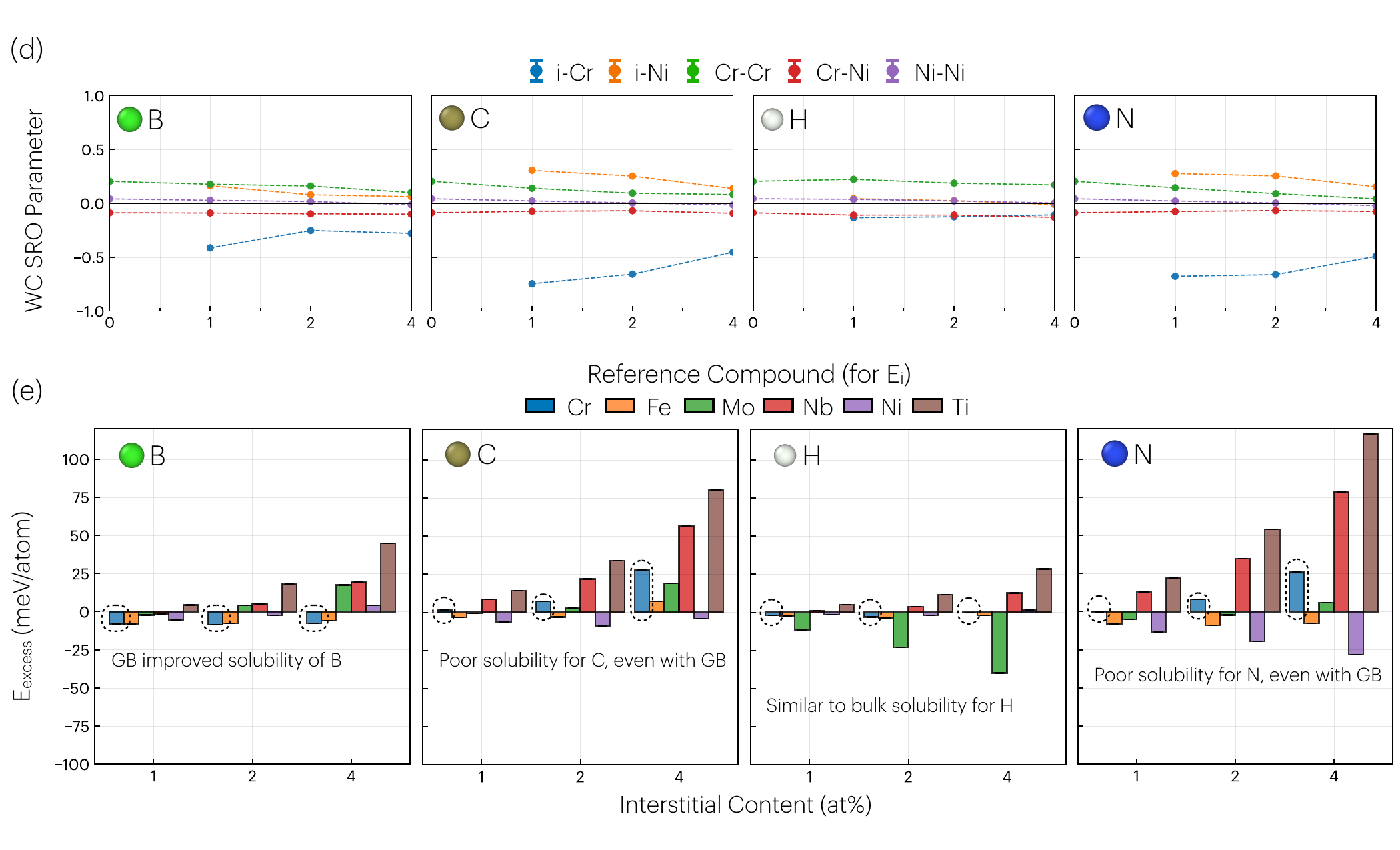}
    \caption{(a) The undoped Cr\textsubscript{30}Ni grain boundary (GB) simulation cell, shown without and with the grain segmentation modification applied in OVITO. (b) The atomic composition of the equilibrated GB simulation cells as a function of the z-axis. Colors distinguish chemical species, while marker symbols indicate different dopant concentrations. (c) The GB energy ($\gamma_{\rm GB}$) and segregation energy ($E_{\rm seg}$) as a function of interstitial content. For better reference, the interstitial content \textit{in the GB zone} is annotated near the corresponding $\gamma_{\rm GB}$ value. (d) The SRO parameter ($\alpha_{ij}$) plotted as a function of interstitial content. (e) The excess energy ($E_{\rm excess}$) plotted as a function of interstitial content and reference compound used for interstitial chemical potential energy ($E_{i}$).}
    \label{fig:gb-energetics}
\end{figure}

\begin{table}[H]
    \centering
    \resizebox{\linewidth}{!}{
    \begin{tabular}{c | c c c | c c c | c c c | c c c}
        \toprule\toprule
        Light & \multicolumn{3}{c|}{$x_{\rm GB}^{(i)}$ (at\%)} 
        & \multicolumn{3}{c|}{$\mu$ (\%)} 
        & \multicolumn{3}{c|}{$\sigma$ (\%)}  
        & \multicolumn{3}{c}{$\tilde{R}_q$ (\AA)} \\

        Interstitial & 1  & 2  & 4  & 1  & 2  & 4  & 1  & 2  & 4 & 1 & 2 & 4 \\
        \midrule
        B & 3.09 & 9.60 & 12.2 & 1.62 & 0.933 & -0.132 & 4.31 & 4.55 & 5.07 & -0.249 & -0.409 & -0.293 \\
        C & 4.00 & 6.77 & 11.8 & 2.14 & 1.52 & 0.922 & 4.55 & 4.51 & 4.68 & -0.133 & -0.249 & -0.197 \\
        H & 3.10 & 4.94 & 10.3 & 1.91 & 1.32 & -1.18 & 4.75 & 4.84 & 6.09 & -0.216 & 0.583 & -0.0486 \\
        N & 4.62 & 6.50 & 9.80 & 2.20 & 1.06 & 1.68 & 4.31 & 4.54 & 4.83 & 0.115 & -0.225 & -0.185 \\ 
        \midrule
        Undoped & \multicolumn{3}{c|}{--} & \multicolumn{3}{c|}{3.20} & \multicolumn{3}{c|}{4.45} & \multicolumn{3}{c}{2.18}\\
        \bottomrule\bottomrule
    \end{tabular}}
    \caption{Atomic fraction of the interstitial in the GB region ($x_{\rm GB}^{(i)}$), mean ($\mu$) and standard deviation ($\sigma$) of the relative free volume ($\tilde{V}$, defined in Eq. \ref{eq:free_volume}) for only the \textit{metallic atoms} (Cr, Ni) in the GB region, as well as the absolute difference in RMS roughness ($\tilde{R}_q$, defined in Eq. \ref{eq:rmsr} and Eq. \ref{eq:abs_diff}) for the equilibrated GB cell at total dopant concentrations of 1, 2, and 4 at\%. The undoped equilibrated GB simulation cell values serve as a baseline reference.}
    \label{tab:free_volume}
\end{table}
The data in Table \ref{tab:free_volume} highlights the effect of increasing interstitial content on the free volume ($V_{\rm GB} - \langle V_{\rm bulk}\rangle$) of metallic atoms in the GB. At 1 at\% doping, all interstitials reduced the GB free volume relative to the undoped system ($\mu = 3.20\%$), with B exhibiting the most significant reduction ($\mu = 1.62\%$, $x_{\rm GB}^{(i)} = 3.09\%$). This was followed by H ($\mu = 1.91\%$, $x_{\rm GB}^{(i)} = 3.10\%$), C ($\mu = 2.14\%$, $x_{\rm GB}^{(i)} = 4.00\%$), then N ($\mu = 2.20\%$, $x_{\rm GB}^{(i)} = 4.62\%$). These values suggest that B was the most effective in densifying the GB and enhancing cohesion, while H exhibited the highest standard deviation ($\sigma = 4.75\%$), indicating that it introduced local structural variability within the GB. The reduction in free volume results from interstitial segregation to the GB, where they occupy previously unoccupied voids. At 2 at\% and 4 at\% doping, interstitials continued to fill the GB, with H showing the largest reduction in free volume and highest structural variability. This corresponded to an increase in $x_{\rm GB}^{(i)}$ for all species, indicating greater interstitial enrichment in the GB region at higher concentrations, with B having the highest interstitial content at the GB. Unlike B and H, which further reduced GB free volume, C and N showed diminishing effects, suggesting that their densification tendencies may saturate beyond a certain concentration. Increased standard deviations at higher interstitial contents, particularly for B and H at 4 at\%, indicate localized structural distortions. The H-doped GB structure at 4 at\% exhibited the most negative free volume ($\mu = -1.18\%$) and highest variability ($\sigma = 6.09\%$). This suggests that at high H concentrations, the GB undergoes pronounced compaction, which may also introduce localized stress concentrations. Such behavior aligns intuitively with H’s small atomic radius, high diffusivity, and balanced chemical affinity for both Cr and Ni in the CrNi lattice. Considering surface roughness, negative values indicate that the doped GB planes were smoother than the undoped GB plane, demonstrating that light interstitials improved GB smoothness as they filled the voids between the metallic constituents, promoting a reduction in the $\gamma_{\rm GB}$. Among the dopants, B produced the flattest planes at all dopant concentrations (most negative values), which corresponded to a continual decrease in $\gamma_{\rm GB}$ with increasing B content (see Fig. \ref{fig:gb-energetics}c). This suggests B improves cohesion among the GB constituents. Notably, there is a correlation between $R_q$ and $\gamma_{\rm GB}$, where rougher surfaces exhibited higher $\gamma_{\rm GB}$. Consider the case for 2 at\% H, where the increased roughness ($\tilde{R_q} = 0.583$) corresponded to a $\gamma_{\rm GB}$ value that exceeded $\gamma^{(\rm undoped)}_{\rm GB}$. This analysis highlights that all interstitials reduced GB free volume and smoothed the GB plane (with 2 at\% H as the exception). Among the light interstitials, B and H were the most effective at compacting the GB with increasing concentration, whereas C and N showed diminishing effects at higher concentrations, suggesting a rapid saturation of their GB stabilization influence.
The most significant findings from the GB simulations are compiled in Fig. \ref{fig:gb-energetics}. To begin, Fig. \ref{fig:gb-energetics}a presents the undoped GB simulation cell as visualized in OVITO with the grain segmentation modification applied. The atomic composition was analyzed layer-by-layer along the z-axis (visualized in Fig. \ref{fig:gb_supp}c) to highlight GB segregation per species and is given in Fig. \ref{fig:gb-energetics}b. In the undoped system, a Cr enrichment peak (orange curve) appears in the central GB region (near 22.5 \AA) and at the periodic boundary condition (PBC) GB (0 and 38 \AA), indicating that Cr preferentially segregated to the GB. This Cr segregation trend was consistently observed in all doped systems. Additionally, in the doped systems interstitial segregation peaks (blue curves) were found around 20~\AA{} in the central GB and at 0 and 38~\AA{} at the PBC GB, confirming that light interstitials preferentially segregated to the GB. The peaks in Cr and interstitial content correspond to troughs in Ni content (green curves), indicating a redistribution of atomic species near the GB. However, in the H-doped cell, the inverse correlation between Cr and Ni is considerably less pronounced. This behavior aligns with the SRO results, which demonstrated that H had an affinity for both Cr and Ni, thereby reducing Ni migration away from the GB. As a result, H promoted Ni enrichment in regions that would otherwise experience Ni depletion. This altered chemical ordering could serve as an early-stage mechanism for GB decohesion. Furthermore, in the undoped and H-doped system, the central GB location migrated upwards from the initial placement around 20~\AA{} to between 21 and 22.5~\AA{}, respectively. This indicates that H had a limited impact on GB migration, despite producing a large impact on GB morphology (from Table \ref{tab:free_volume}). In the C- and N-doped systems, no significant migration occurred, indicating the formation of more rigid clusters. Meanwhile, in the B-doped system, there was a small migration downwards, towards 19~\AA{}, as a result of B's ability to produce a smoother GB plane, effectively `flattening' the GB structure.

The energetics of the GB simulation cells are presented in Fig. \ref{fig:gb-energetics}c, where the top figure shows $\gamma_{\rm GB}$ as a function of interstitial content and the bottom figure depicts $E_{\rm seg}$ versus interstitial content. Due to the symmetry in the way these values are calculated, similar trends appear in both plots, though they offer distinct insights. For B, C, and N, there was reduction in $\gamma_{\rm GB}$ relative to the undoped value of 0.800 J/m\textsuperscript{2}, suggesting that the segregation of these light elements into previously unoccupied voids stabilized the overall GB structure. The magnitude of this stabilization varied depending on the species, with B exhibiting the strongest stabilizing effect, as it continually decreased $\gamma_{\rm GB}$ as it filled the GB zone. Boron also had the lowest segregation energy, indicating that it is the most effective GB segregator in terms of thermal stability and the most likely to preferentially segregate among the interstitials considered. Unlike the other elements, B's impact on segregation and stabilization did not saturate up to 4 at\% total dopant concentration; the GB continued to stabilize as increasing amounts of B occupied the boundary region. Following B, C exhibited stabilization and segregation effects that quickly saturated beyond 2 at\%, indicating a critical trade-off in C segregation at the GB. Excessive C content could potentially destabilize the GB in the absence of alternate segregation pathways, such as carbide precipitation. Nitrogen showed an even faster saturation than C, with a noticeable increase in $\gamma_{\rm GB}$ and $E_{\rm seg}$ from 1 at\% to 4 at\%. This suggests that excessive N accumulation in the GB zone would eventually compromise the structural stability of the GB. Lastly, H exhibited relatively consistent behavior from 1 at\% to 4 at\%, with a notable increase in $\gamma_{\rm GB}$ at 2 at\%. This increase is attributed to reduced cohesion among metallic constituents, likely driven by increased structural variability and GB surface roughness (see Table \ref{tab:free_volume}). Since the MC routine equilibrated the system to its thermodynamically preferred state, the high concentration of H within the GB zone confirms its strong preference for the GB environment. However, despite this preference, the energetic benefit of H segregation remained minimal or deleterious, resulting in geometric disorder, which is likely to drive the system toward a state prone to embrittlement.

The SRO results for the GB simulation cell are presented in Fig. \ref{fig:gb-energetics}d. Co-segregation of Cr and light interstitials to the GB region resulted in negative $\alpha_{\rm iCr}$ values, indicating a strong local preference for Cr. Similar to the bulk system, C and N exhibited the strongest affinity for Cr, further enriching the GB region with Cr as they segregated. This is reflected in the lower $\alpha_{\rm CrCr}$ values with increased C and N content. Additionally, the M-M' $\alpha_{ij}$ values for Cr-Ni and Cr-Cr decreased in the undoped GB system relative to the undoped pristine system, suggesting that the GB environment altered the preferred atomic ordering. These insights highlight the dual influence of chemical composition and geometric constraints in shaping SRO behavior, demonstrating that atomic ordering is governed not only by elemental interactions but also by the spatial configuration offered in the GB.

The excess energy results are presented in Fig. \ref{fig:gb-energetics}e, showing a small increase in the number of negative values compared to those reported for bulk Cr\textsubscript{30}Ni. Combined with the bulk predictions, these results provide a more comprehensive understanding of the segregation and precipitation tendencies of light interstitials. For example, B previously exhibited a strong tendency to leave the Cr\textsubscript{30}Ni lattice (see Fig. \ref{fig:excess_energy}b). However, when given the opportunity to segregate to the GB, B no longer exhibits an energetic preference to ‘escape’ the simulation cell. Even as more B accumulated within the GB zone, its B-Cr excess energy remained largely unchanged, indicating strong solubility of B in the Cr-Ni GB. Notably, the likelihood of B segregation increased in the presence of Mo, Nb, and Ti. This suggests that the chemical affinity of B for these elements, combined with its GB solubility, may lead to the enrichment of Mo, Nb, and/or Ti around B-doped GBs. Together with the bulk insights, the excess energy and GB energetics confirm that B not only exhibits a thermodynamic preference to leave the bulk lattice but also demonstrates a strong segregation tendency toward GBs. Among all the light interstitials considered in this study, B shows the highest GB enrichment and the most pronounced stabilization effect. This behavior suggests that B does not favor the formation of precipitates or secondary phases within the microstructure; instead, it preferentially segregates to GBs, where it plays a critical role in enhancing cohesion and reducing the GB energy. As such, B can be classified as the most potent GB stabilizer among the light interstitials evaluated here.

Meanwhile, C and N exhibit distinct behavior, showing a preference to associate with Cr rather than remain uniformly distributed within the Cr\textsubscript{30}Ni GB simulation cell. This prediction aligns well with established experimental observations that C and N tend to form stable carbides or nitrides, rather than remain dissolved within the GB region \cite{bianGrainBoundaryDiffusion2024, weiMicrostructureEvolutionCreeprupture2023, liuEffectsNbAddition2024, smithRoleNiobiumWrought2005}. As the concentrations of C and N increased within the GB zone, the probability of Cr–(C,N) association also increased, reinforcing the likelihood of Cr-based carbide and nitride formation under such conditions. For H, its segregation behavior showed minimal deviation from the bulk $E_{\rm excess}$ values. However, the MC routine confirmed that H aggregation at the GB is thermodynamically preferred. Since H is typically introduced into the material through environmental exposure, GBs would have already formed prior to significant H ingress. These findings suggest that H preferentially migrates along GB channels rich in Cr and Ni, while the increasingly negative $E_{\rm excess}$ values for H-Mo indicate a strong tendency for H to avoid Mo-enriched GBs. This is consistent with the repulsive interactions observed in both the GB and bulk environments (see Fig. \ref{fig:excess_energy}b) and aligns with recent studies highlighting Mo’s role in mitigating HE \cite{leeComparativeStudyEffects2019, leeRoleMoCarbides2015, leeEffectsVanadiumCarbides2016, liuEffectsMoConcentration2024}. Alternatively, increased Nb or Ti concentrations at the GB appear to facilitate H diffusion, though to a lesser extent than Mo’s repulsive effect. This segregation behavior could have important implications for HE mechanisms, particularly in service environments where H ingress is a concern.

\section{Conclusion}
This study systematically investigated the behavior of light interstitials (B, C, H, and N) in a Cr\textsubscript{30}Ni alloy, considering both bulk and grain boundary (GB) environments. Through Monte Carlo simulations, we identified clear chemical and geometric preferences for each interstitial, providing atomistic insights corroborated to experimentally observed segregation and precipitation trends in Ni-based superalloys.

A key finding of this study was boron's strong preference to leave the bulk lattice and segregate to GBs. Once in the GBs, B had the most pronounced effect, where it reduced the GB energy and promoted a more cohesive GB. This aligned well with experimental observations that report B’s role in enhancing GB cohesion which promotes exceptional creep resistance at high temperatures. Interestingly, B was the only light interstitial to induce a structural transformation in the bulk, disrupting the original ordering and indicating a strong driving force for segregation. Upon removing B and re-relaxing the system, the metal lattice largely reverted to its primary FCC structure. This behavior aligns with B’s experimentally observed role in promoting serrated GBs, which are known to enhance resistance to GB sliding and improve high-temperature mechanical properties. These atomistic insights reinforce the well-established understanding that B primarily acts as a GB strengthening element rather than driving bulk precipitation. In contrast, carbon and nitrogen exhibited strong ordering with Cr in both bulk and GB environments, forming carbide- and nitride-like motifs consistent with experimentally observed precipitation behavior in high-temperature alloys. Additionally, their solubility within the Cr-Ni GB was limited, instead preferring to segregate with Cr over remaining dissolved in the GB. This aligns with experimental observations of carbide and nitride precipitation rather than a homogeneous distribution. Importantly, the shared chemical affinities and GB segregation tendencies of B, C, and N highlight the potential for complex multi-element interactions, including the formation of mixed boro-nitro-carbides, borocarbides, and carbonitrides. These interactions could further influence microstructural stability and mechanical performance.

Hydrogen exhibited strong aggregation within the GB structure, suggesting that Cr-Ni GBs may serve as diffusion pathways for H ingress. Unlike other light interstitials, H segregation did not enhance GB stability but instead introduced additional atomic disorder. This finding aligns with experimental studies on hydrogen embrittlement in Ni-based alloys, where H-induced decohesion frequently occurs along GBs. Additionally, H was found to avoid ordering with Mo, supporting recent findings on Mo’s role in mitigating embrittlement. The repulsion of H from Mo-rich regions in both bulk and GB environments further reinforces mechanistic explanations of Mo segregation enhancing embrittlement resistance.

Based on our findings, we propose the following hypothesis to the community for future analysis (also illustrated in Fig. \ref{fig:overview}):
\begin{itemize}
\item Mo-containing borides act as effective barriers against hydrogen-induced degradation by inhibiting H ingress and stabilizing GB cohesion.
\end{itemize}

By elucidating both the chemical and structural preferences of these interstitials—specifically their metallic ordering and bulk vs. GB occupancy—our work provides a robust computational framework for interpreting experimental segregation studies. This level of predictive insight is critical for designing next-generation superalloys with improved GB stability, enhanced resistance to embrittlement, and optimized precipitation behavior. Future work will extend these findings by incorporating additional alloying elements and exploring kinetic effects on interstitial diffusion and clustering.

\section*{Author Contributions}
\textbf{T.D.D} Writing - original draft, writing – review \& editing, visualization, validation, software, methodology, investigation, formal analysis, data curation. \textbf{R.F.} and \textbf{J. Li} Project administration and supervision, writing - review \& editing. All authors contributed to the conceptualization of this project.

\section*{Data Availability}
The hybrid Monte Carlo Molecular Dynamics (hMCMD) routine will be made available at \url{https://github.com/tylerdolezal/hybrid_MCMD}. All data generated from this work and post-processing scripts will be provided at \url{https://github.com/tylerdolezal}.

\section*{Acknowledgments}
J. Li acknowledges support from NSF CMMI-1922206 and DMR-1923976.

\bibliographystyle{ieeetr}

\renewcommand{\thefigure}{S\arabic{figure}}
\setcounter{figure}{0}  

\renewcommand{\thetable}{S\arabic{table}}
\setcounter{table}{0}  

\renewcommand{\theequation}{S-\arabic{equation}}
\setcounter{equation}{0}  

\section*{Supplemental Materials}
\subsection*{Excess Energy Reference Compounds}
\begin{table}[H]
    \centering
    \renewcommand{\arraystretch}{1.0} 
    \caption{Complete dataset of reference compounds and computed interstitial chemical potential energies (\(E_i\)) for B and C with Cr, Fe, Mo, Nb, Ni, and Ti.}
    \label{tab:supplemental_materials_energies_1}
    \resizebox{\linewidth}{!}{ 
    \begin{tabular}{c c c}
        \toprule\toprule
        & \textbf{Boron (B)} & \textbf{Carbon (C)} \\
        \midrule
        \multirow{5}{*}{\rotatebox{90}{}}
        & \begin{tabular}{l l c}
            Compound & Space Group & \(E_i\) (eV/atom) \\
            \midrule
            CrB   & I4$_1$/amd & -7.619 \\
            CrB$_4$   & Pnnm & -6.874 \\
            CrB$_4$   & Immm & -6.857 \\
            CrB   & Cmcm & -7.613 \\
            CrB$_2$   & P6/mmm & -7.020 \\
            Cr$_2$B   & Fddd & -7.745 \\
            Cr$_2$B   & I4/mcm & -7.736 \\
            Cr$_5$B$_3$   & I4/mcm & -7.725 \\
            Cr$_3$B$_4$   & Immm & -7.367 \\
            Cr$_2$B$_3$   & Cmcm & -7.254 \\
            FeB   & Cmcm & -7.313 \\
            FeB$_2$   & Pnma & -6.914 \\
            FeB   & I4$_1$/amd & -7.306 \\
            FeB   & Pnma & -7.329 \\
            FeB   & I4/mcm & -7.312 \\
            Fe$_2$B   & I4/mcm & -7.554 \\
            Fe$_3$B   & Pnma & -7.488 \\
            Fe$_3$B   & I-4 & -7.440 \\
            Fe$_{23}$B$_6$   & Fm$\bar{3}$m & -7.412 \\
            FeB$_4$   & Pnnm & -6.716 \\
            Fe$_2$B$_7$   & Pbam & -6.693 \\
            MoB   & I4$_1$/amd & -7.525 \\
            MoB$_2$   & R-3m & -7.148 \\
            MoB$_3$   & R-3m & -6.852 \\
            Mo$_2$B   & I4/mcm & -7.451 \\
            Mo$_3$B$_2$   & P4/mbm & -7.481 \\
            NbB   & Cmcm & -8.054 \\
            NbB$_2$   & P6/mmm & -7.576 \\
            Nb$_3$B$_4$   & Immm & -7.855 \\
            Nb$_2$B$_3$   & Cmcm & -7.761 \\
            Nb$_3$B$_2$   & P4/mbm & -8.087 \\
            Nb$_5$B$_6$   & Cmmm & -7.922 \\
            NiB   & Cmcm & -7.026 \\
            Ni$_4$B$_3$   & Pnma & -7.178 \\
            Ni$_3$B   & Pnma & -7.584 \\
            Ni$_2$B   & I4/mcm & -7.433 \\
            Ni$_4$B$_3$   & C2/c & -7.183 \\
            Ni$_{23}$B$_6$   & Fm$\bar{3}$m & -7.486 \\
            TiB$_2$   & P6/mmm & -8.159 \\
            TiB   & Pnma & -7.635 \\
            Ti$_3$B$_4$   & Immm & -8.234 \\
        \end{tabular}
        & \begin{tabular}{l l c}
            Compound & Space Group & \(E_i\) (eV/atom) \\
            \midrule
            Cr$_{23}$C$_6$   & Fm$\bar{3}$m & -8.604 \\
            Cr$_3$C$_2$   & Pnma & -8.452 \\
            Cr$_7$C$_3$   & P6$_3$mc & -8.545 \\
            Cr$_7$C$_3$   & Pnma & -8.551 \\
            Cr$_3$C   & Pnma & -8.550 \\
            Cr$_2$C   & P$\bar{3}$m1 & -8.218 \\
            CrC   & P$\bar{6}$m2 & -8.150 \\
            Fe$_{23}$C$_6$   & Fm$\bar{3}$m & -8.073 \\
            Fe$_3$C   & P6$_3$22 & -7.917 \\
            Fe$_3$C   & Pnma & -8.014 \\
            Fe$_5$C$_2$   & C2/c & -8.022 \\
            Fe$_2$C   & Pnnm & -7.982 \\
            Fe$_5$C$_2$   & P2/c & -8.018 \\
            Fe$_7$C$_3$   & Pnma & -7.982 \\
            Fe$_7$C$_3$   & P6$_3$mc & -7.952 \\
            Mo$_2$C   & Pbcn & -8.380 \\
            MoC   & P$\bar{6}$m2 & -8.296 \\
            Nb$_2$C   & Pnma & -9.354 \\
            NbC   & Fm$\bar{3}$m & -8.994 \\
            Ni$_3$C   & R$\bar{3}$c & -7.776 \\
            Ti$_2$C   & Fd$\bar{3}$m & -9.963 \\
            TiC   & Fm$\bar{3}$m & -9.775 \\
            &&\\&&\\&&\\&&\\&&\\&&\\&&\\&&\\
            &&\\&&\\&&\\&&\\&&\\&&\\&&\\&&\\
            &&\\&&\\&&\\
        \end{tabular} \\
        \bottomrule\bottomrule
    \end{tabular}
    }
\end{table}

\begin{table}[H]
    \centering
    \renewcommand{\arraystretch}{1.0} 
    \caption{Complete dataset of reference compounds and computed interstitial chemical potential energies (\(E_i\)) for H and N with Cr, Fe, Mo, Nb, Ni, and Ti.}
    \label{tab:supplemental_materials_energies_2}
    \resizebox{\linewidth}{!}{ 
    \begin{tabular}{c c c}
        \toprule\toprule
        & \textbf{Hydrogen (H)} & \textbf{Nitrogen (N)} \\
        \midrule
        \multirow{5}{*}{\rotatebox{90}{}}
        & \begin{tabular}{l l c}
            Compound & Space Group & \(E_i\) (eV/atom) \\
            \midrule
            CrH   & Fm$\bar{3}$m & -2.380 \\
            CrH$_2$   & Fm$\bar{3}$m & -2.366 \\
            FeH   & Fm$\bar{3}$m & -2.339 \\
            MoH   & P6$_3$/mmc & -1.355 \\
            NbH$_2$   & Fm$\bar{3}$m & -2.720 \\
            NbH   & Cmme & -2.651 \\
            NbH   & I4/mmm & -2.649 \\
            NiH   & Fm$\bar{3}$m & -2.434 \\
            Ni$_2$H   & R$\bar{3}$m & -2.418 \\
            Ni$_2$H   & P$\bar{3}$m1 & -2.422 \\
            TiH$_2$   & I4/mmm & -3.127 \\
            TiH$_2$   & Fm$\bar{3}$m & -3.132 \\
            TiH   & P4$_2$/mmc & -3.092 \\
            Ti$_4$H$_7$   & P$\bar{4}$3m & -3.115 \\
            Ti$_2$H$_3$   & P4$_2$/mcm & -3.093 \\
            Ti$_4$H$_5$   & P$\bar{4}$2m & -3.086 \\
            Ti$_4$H$_3$   & P$\bar{4}$2m & -3.015 \\
            TiH   & Fm$\bar{3}$m & -3.124 \\
            Ti$_2$H   & Pn$\bar{3}$m & -2.998 \\
            TiH$_2$   & P4/nmm & -3.021 \\
            &&\\&&\\&&\\&&\\&&\\
        \end{tabular}
        & \begin{tabular}{l l c}
            Compound & Space Group & \(E_i\) (eV/atom) \\
            \midrule
            Cr$_2$N   & P$\bar{3}$1m & -6.645 \\
            CrN   & Pmmn & -6.118 \\
            CrN   & P6$\bar{3}$m2 & -6.476 \\
            CrN   & F$\bar{4}$3m & -6.215 \\
            Cr$_3$N$_2$   & R$\bar{3}$c & -6.425 \\
            Cr$_3$N$_2$   & C2/m & -6.394 \\
            Cr$_3$N$_4$   & P6$_3$/m & -5.846 \\
            Fe$_3$N   & P6$_3$22 & -5.769 \\
            FeN   & F$\bar{4}$3m & -5.692 \\
            Fe$_8$N   & I4/mmm & -5.671 \\
            Fe$_2$N   & Pbcn & -5.641 \\
            Fe$_4$N   & Pm$\bar{3}$m & -5.589 \\
            MoN   & P6$_3$mc & -6.122 \\
            Mo$_2$N   & I4$\bar{1}$/amd & -6.246 \\
            MoN   & P6$_3$/mmc & -6.216 \\
            MoN   & P$\bar{6}$m2 & -6.275 \\
            Nb$_2$N   & P$\bar{3}$1m & -8.007 \\
            NbN   & P6$_3$/mmc & -7.472 \\
            Nb$_5$N$_6$   & P6$_3$/mcm & -7.218 \\
            Nb$_2$N$_3$   & Pnma & -6.819 \\
            Ni$_3$N   & P6$_3$22 & -5.241 \\
            TiN   & Fm$\bar{3}$m & -9.004 \\
            Ti$_2$N   & P4$_2$/mnm & -9.509 \\
            Ti$_3$N$_2$   & R$\bar{3}$c & -9.270 \\
            Ti$_8$N$_5$   & P$\bar{4}$m2 & -9.270 \\
        \end{tabular} \\
        \bottomrule\bottomrule
    \end{tabular}
    }
\end{table}

\begin{table}[H]
\centering
\caption{Interstitial chemical potential energies derived from DFT and Matlantis, based on total energies of selected binary compounds. Values are given in eV per interstitial atom.}
\begin{tabular}{lccc}
\toprule\toprule
\textbf{Atomic Pair} & \textbf{MP ID} & \textbf{DFT $\mu_i$ (eV/atom)} & \textbf{Matlantis $\mu_i$ (eV/atom)} \\
\hline
Cr--B & mp-1080664 & $-7.738$ & $-7.419$ \\
Mo--B & mp-1890    & $-7.684$ & $-7.525$ \\
Ti--B & mp-1145    & $-8.263$ & $-8.254$ \\
Cr--C & mp-723     & $-9.573$ & $-8.604$ \\
Mo--C & mp-1552    & $-9.566$ & $-8.380$ \\
Ti--C & mp-10721   & $-11.158$ & $-9.963$ \\
Cr--H & mp-24208   & $-3.452$ & $-2.380$ \\
Mo--H & mp-24417   & $-2.426$ & $-1.355$ \\
Ti--H & mp-24726   & $-4.118$ & $-3.127$ \\
Cr--N & mp-8780    & $-9.494$ & $-6.118$ \\
Mo--N & mp-2811    & $-9.070$ & $-6.122$ \\
Ti--N & mp-492     & $-11.770$ & $-9.509$ \\
\bottomrule\bottomrule
\end{tabular}
\label{tab:mu_comparison}
\end{table}

Table \ref{tab:mu_comparison} presents a side-by-side comparison of interstitial chemical potential energies derived from DFT and Matlantis calculations, using binary compounds as reference structures. Across all atomic pairs, Matlantis predictions are consistently less negative than their DFT counterparts, with deviations ranging from 0.01-2.2 eV. Despite this offset, Matlantis preserves the relative ranking and chemical affinity trends across different host metals, validating its use as a thermodynamically consistent surrogate model. Notably, strong affinities for Ti are retained, as is the significant reduction in affinity for Mo-H, all of which support the chemical ordering trends captured in our excess energy framework.

\subsection*{Constructing Isotherms}
Traditional bulk-to-GB segregation models rely on isotherm-based approaches \cite{tuchindaGrainSizeDependencies2022, hofmannSoluteSegregationGrain1996}, such as the McLean isotherm \cite{mcleanGrainBoundariesMetals1957}, which describe solute occupancy as a function of bulk concentration under equilibrium conditions. However, these models often average over site-specific variations and do not explicitly resolve the atomic-scale interactions that drive segregation and structural transformations \cite{lejcekGrainBoundarySegregation2010}. To overcome these limitations, this work adopts an alternative \(E_{\text{excess}}\)-based framework, which quantifies segregation energetics relative to different segregation routes, much like how surface adsorption coverage depends on the partial pressure of an adsorbate (e.g., $P_{\rm O_2}$ in oxidation studies). By first equilibrating the system through MC simulations, this approach captures site preferences, clustering behavior, and structural shifts arising from interstitial content. The equilibrated configurations are then analyzed using the \(E_{\text{excess}}\) method, which compares the system's energy against various reference compounds to quantify the thermodynamic driving forces for segregation. This combined MC-\(E_{\text{excess}}\) framework provides insight into how metallic constituents promote interstitial segregation, influencing GB enrichment and secondary phase formation. Unlike traditional isotherms, which average over atomic configurations, this method offers an atomistically resolved segregation description, making it well-suited for analyzing multi-element interactions.

Beyond characterizing segregation energetics, \(E_{\text{excess}}\) can be leveraged to construct an isotherm formulation that quantitatively predicts solute retention in the Cr\textsubscript{30}Ni bulk phase. This approach captures both the energetic preference for segregation and the competition between enthalpy-driven ordering and entropy-driven bulk solubility at elevated temperatures. To formalize this, the McLean isotherm equation \cite{mcleanGrainBoundariesMetals1957} is introduced where the solute fraction remaining in the bulk structure follows a Boltzmann-like dependence on \(E_{\text{excess}}\), given by Eq. \ref{eq:isotherm}:

\begin{equation}\label{eq:isotherm}
\tilde{X} = \frac{X^{(i)} \exp\left(-\frac{\Delta \tilde{G}}{k_{\rm B}T} \right)}
{1 + X^{(i)} \left( \exp\left(-\frac{\Delta \tilde{G}}{k_{\rm B}T} \right) - 1 \right)} \text{ ,}
\end{equation}
where \(\tilde{X}\) represents the effective light interstitial concentration remaining in the bulk Cr\textsubscript{30}Ni matrix after accounting for segregation tendencies, and \(X^{(i)}\) is the total interstitial content before considering segregation effects. For clarification, \(X^{(i)}\) represents the actual interstitial concentration in the bulk system, while \(\tilde{X}\) is the predicted interstitial concentration after accounting for segregation tendencies relative to the reference compound. The term \(\Delta \tilde{G}\) quantifies the energetic preference for segregation and is given in Eq. \ref{eq:delta_G}:

\begin{equation}\label{eq:delta_G}
    \Delta \tilde{G} = E_{\text{excess}} - T\Delta S_{\text{config}} \text{ ,}
\end{equation}
where the configurational entropy (\(\Delta S_{\text{config}}\)) is approximated using the ideal mixing expression, \(\Delta S_{\text{config}} = -k_B\sum x_i \ln x_i\). Note that in the limit \( k_B T \gg \Delta \tilde{G} \), Eq. \ref{eq:isotherm} reduces to \(\tilde{X} \approx X^{(i)}\). Physically, this corresponds to a regime where entropy-driven mixing overcomes the enthalpic driving force for segregation. This behavior is consistent with the expectation that at elevated temperatures, \(\Delta S_{\text{config}}\) favors a more randomized atomic arrangement. This formulation provides a quantitative prediction for the solubility of interstitial species in the Cr\textsubscript{30}Ni bulk phase as a function of bulk composition, segregation energetics, and temperature. By incorporating multiple reference compounds, this approach effectively models competing segregation pathways and their influence on bulk solubility (displayed in Fig. \ref{fig:bulk_isotherms}).

For GB structures, the \(E_{\text{excess}}\) term in Eq. \ref{eq:delta_G} can be replaced with \(E_{\text{seg}}\), where \(E_{\text{seg}}\) quantifies the segregation energy relative to the bulk (defined in Eq. \ref{eq:segregation}). In this case, \(\tilde{X}\) represents the predicted light interstitial concentration at the GB. Ultimately, this approach extends beyond the limitations of classical segregation models \cite{lejcekGrainBoundarySegregation2010} by incorporating explicit energy-based site preference and atomic ordering effects. Unlike traditional isotherm formulations, which assume uniform site distributions, this framework integrates GB segregation energy within a temperature-dependent isotherm, providing a direct thermodynamic measure of solute stability at the GB. By leveraging segregation energy alongside excess energy referenced to stable compounds, this model captures the interplay between bulk solubility, GB enrichment, and phase stability across varying thermal conditions. This refined approach enables a more detailed understanding of segregation-driven transformations in complex alloy systems.
\subsection*{Additional Results}
\begin{figure}[H]
    \centering
    \includegraphics[width=\linewidth]{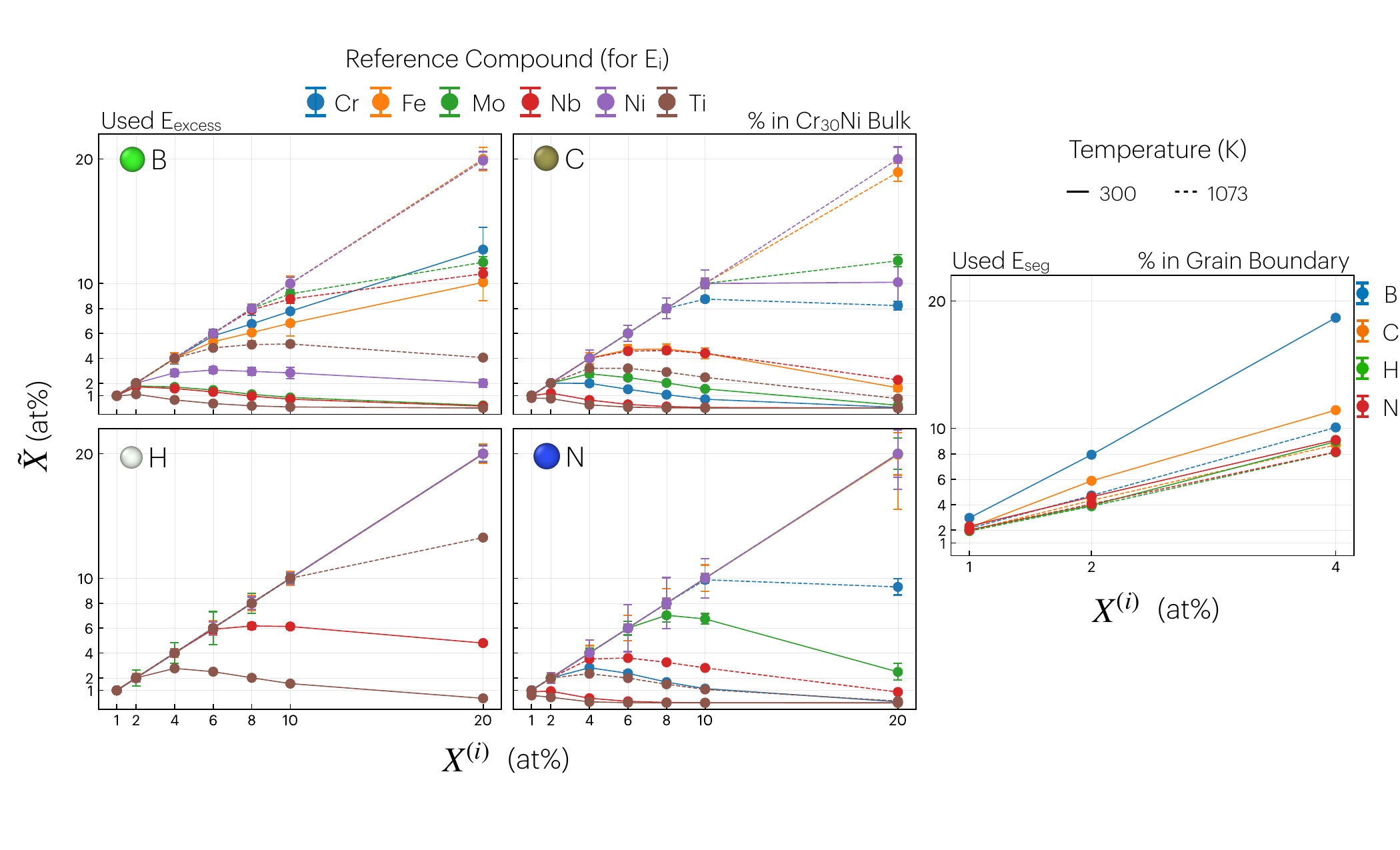}
    \caption{(a) The predicted interstitial concentration (\(\tilde{X}\)), as defined in Eq. \ref{eq:isotherm}, is plotted against the total interstitial content for each interstitial type. The color scheme corresponds to the reference compound used to compute \(E_{i}\), as given in Eq. \ref{eq:mu}. The label corresponds to the metal component of each reference compound, e.g., Cr for Cr-i (i = B, C, H, N) or Fe for Fe-i (i = B, C, H, N), etc. (b) The predicted interstitial concentration in the grain boundary (\(\tilde{X}\)), as defined in Eq. \ref{eq:isotherm} with the exception that \(E_{\rm seg}\) takes the place of \(E_{\rm excess}\) in \(\Delta \tilde{G}\), plotted against total interstitial content for each interstitial type. For both diagrams, the line style differentiates temperature: solid lines represent 300 K, and dashed lines represent 1073 K.}
    \label{fig:bulk_isotherms}
\end{figure}

The isotherms are compiled in Fig. \ref{fig:bulk_isotherms}, with the left panel presenting results for the bulk Cr\textsubscript{30}Ni system and the right panel showing a single graph for the GB system. For the left panel, to interpret these curves, an upward trend indicates that more of the light interstitial remains within the Cr\textsubscript{30}Ni matrix, while a downward trend suggests that the interstitial prefers to leave the matrix and form ordering with a reference metallic species (Cr, Fe, Mo, Nb, Ni, or Ti). For instance, at both 300 K and 1073 K, B exhibits a strong preference for ordering with Ti (brown solid and dashed curves). At 300 K, B also shows strong segregation with Ni (solid purple line), whereas at 1073 K, it remains dissolved in the Cr\textsubscript{30}Ni matrix (dashed purple line). These trends illustrate the segregation tendencies at both low and high temperatures, highlighting the entropic counterforce that drives a more random solid solution at elevated temperatures. Most notably, dashed lines that remain low on the y-axis indicate strong thermal stability of the interstitial-metal interactions. For the GB isotherms, \(E_{\rm seg}\) was used in Eq. \ref{eq:delta_G}, where the y-axis represents the fraction of the light interstitial within the GB (with \(1 - y\) corresponding to the fraction outside the GB). Among all interstitials, B exhibited the strongest segregation tendency, with a GB concentration of approximately 18\% within the GB at a total dopant concentration of 4 at\%. As temperature increases, entropy drives the system toward a more random solid solution, leading to a reduction in predicted GB segregation. For comparison, the MC results equilibrated to a GB concentration of 11\% B within the GB.

\begin{figure}[H]
    \centering
    \includegraphics[width=\linewidth]{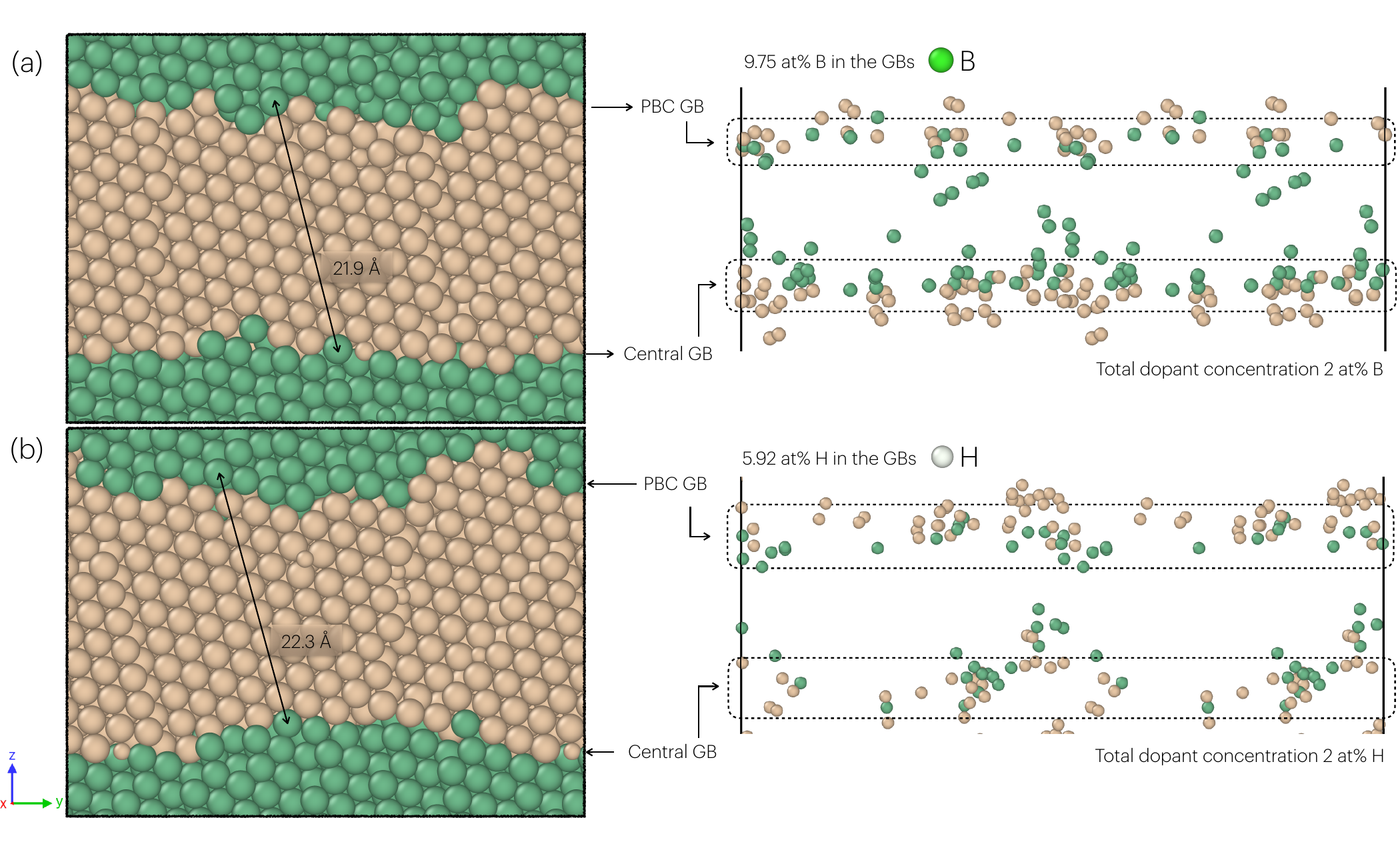}
    \includegraphics[width=\linewidth]{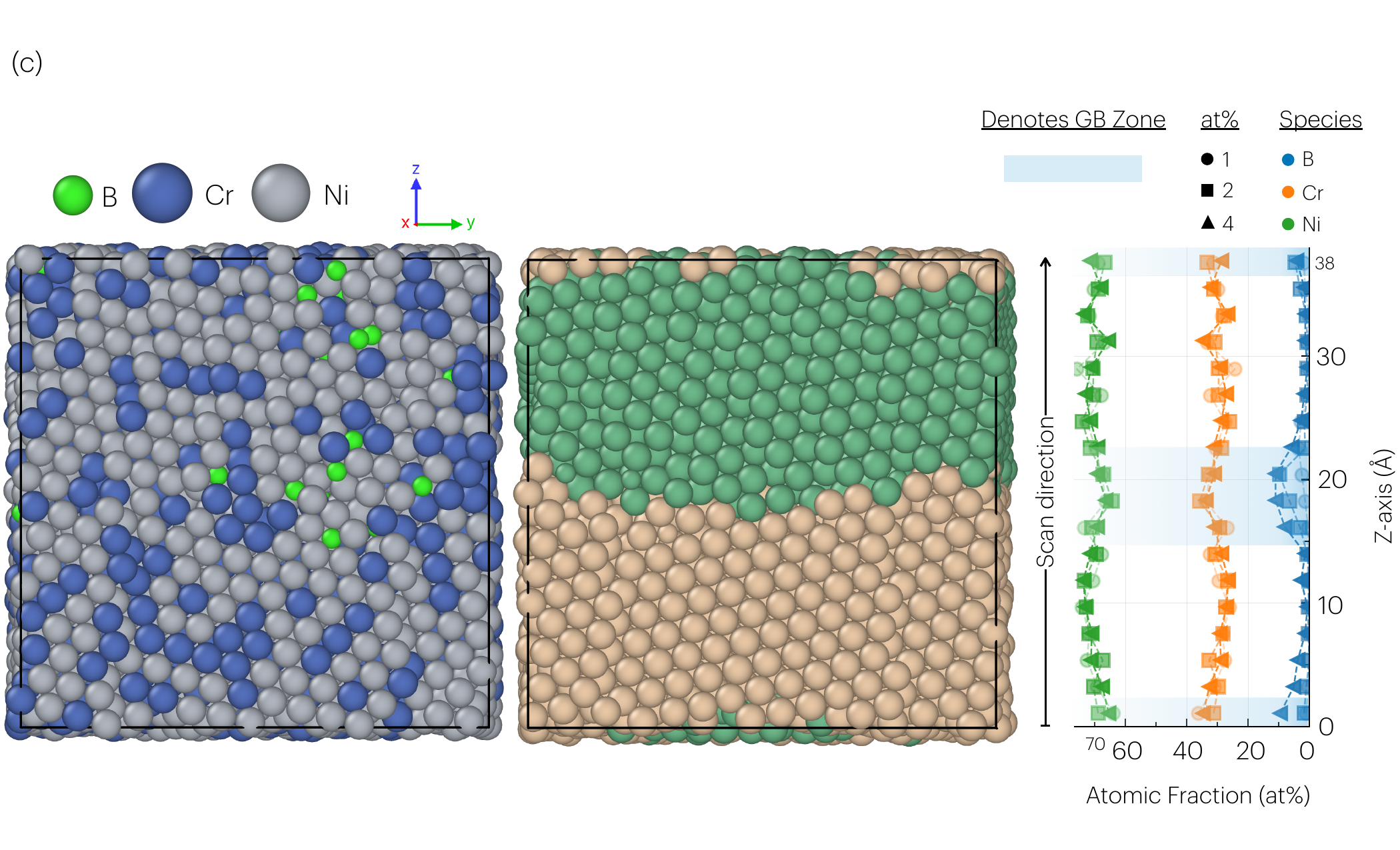}
    \caption{The grain boundary (GB) structure of the (a) B-doped and (b) H-doped simulation cells, visualized in OVITO with the grain segment modifier activated. Alongside these magnified views, a replicated section of the GB region highlights the distribution of B and H atoms, with metal atoms removed to clearly show interstitial positions. Gold and green represent grains 1 and 2, respectively. (c) A visualization of the z-axis atomic composition scan.}
    \label{fig:gb_supp}
\end{figure}

A detailed look at the GB structure is offered in Fig. \ref{fig:gb_supp}, where the B-doped GB cell is shown in panel (a), the H-doped GB cell is shown in panel (b), and panel (c) presents a graphical representation of the z-axis composition scan, complementing Fig. \ref{fig:gb-energetics}b.

\end{document}